\documentclass[usenatbib]{mn2e}

\pdfoutput=1
\usepackage{natbib}
\usepackage{fixltx2e}
\usepackage{mathptmx}
\usepackage{amsmath}
\usepackage{wasysym}
\usepackage{url}
\usepackage{times}
\usepackage{captcont}
\usepackage{ulem}
\usepackage[pdftex]{graphicx}

\newcommand{\Mpc}{\rm\; Mpc}
\newcommand{\kpc}{\rm\; kpc}

\newcommand{\km}{\rm\; km}

\newcommand{\cm}{\rm\; cm}

\newcommand{\pix}{\rm\; pixel}

\newcommand{\ppix}{\hbox{$\pix^{-1}\,$}}

\newcommand{\yr}{\rm\; yr}
\newcommand{\Gyr}{\rm\; Gyr}

\newcommand{\s}{\rm\; s}

\newcommand{\ks}{\rm\; ks}

\newcommand{\GHz}{\rm\; GHz}

\newcommand{\K}{\rm\; K}

\newcommand{\keVpcmcu}{\hbox{$\keV\cm^{-3}\,$}}

\newcommand{\Msun}{\hbox{$\rm\thinspace M_{\odot}$}}

\newcommand{\Msunpyr}{\hbox{$\Msun\yr^{-1}\,$}}

\newcommand{\keV}{\rm\; keV}

\newcommand{\erg}{\rm\; erg}

\newcommand{\ergps}{\hbox{$\erg\s^{-1}\,$}}

\newcommand{\expmapcorr}{\hbox{$\rm\thinspace photons\pcmsq\ps\ppix$}}

\newcommand{\kmps}{\hbox{$\km\s^{-1}\,$}}

\newcommand{\kmpspMpc}{\hbox{$\kmps\Mpc^{-1}\,$}}

\newcommand{\Zsun}{\hbox{$\thinspace \mathrm{Z}_{\odot}$}}

\newcommand{\amin}{\rm\; arcmin}

\newcommand{\asec}{\rm\; arcsec}

\newcommand{\emm}{\hbox{$\cm^{-5}\,$}}
\newcommand{\sqremm}{\hbox{$\cm^{-5/2}\,$}}
\newcommand{\empasecsq}{\hbox{$\emm\asec^{-2}\,$}}

\newcommand{\pseudoP}{\hbox{$\keV\sqremm\asec^{-2}\,$}}

\newcommand{\pcmsq}{\hbox{$\cm^{-2}\,$}}
\newcommand{\pcmcu}{\hbox{$\cm^{-3}\,$}}
\newcommand{\pcmcus}{\hbox{$\pcmcu\s\,$}}

\newcommand{\ps}{\hbox{$\s^{-1}\,$}}

\begin{document}

\title[The merging cluster Abell 2146]{Shock fronts, electron-ion equilibration and ICM transport processes in the merging cluster Abell 2146}
\author[H.R. Russell et al.]  
    {\parbox[]{7.in}{H.~R. Russell$^1$\thanks{E-mail: 
          helen.russell@uwaterloo.ca}, B.~R. McNamara$^{1,2,3}$, 
        J.~S. Sanders$^4$, A.~C. Fabian$^4$, P.~E.~J. Nulsen$^3$, R.~E.~A. Canning$^4$, S.~A. Baum$^{5,6}$, M. Donahue$^7$, A. Edge$^8$, L.~J. King$^{4,9}$, C.~P. O'Dea$^{3,10}$ \\ 
    \footnotesize 
    $^1$ Department of Physics and Astronomy, University of Waterloo, Waterloo, ON N2L 3G1, Canada\\ 
    $^2$ Perimeter Institute for Theoretical Physics, Waterloo, Canada\\ 
    $^3$ Harvard-Smithsonian Center for Astrophysics, 60 Garden Street, Cambridge, MA 02138, USA\\ 
    $^4$ Institute of Astronomy, Madingley Road, Cambridge CB3 0HA\\
    $^5$ Center for Imaging Science, Rochester Institute of
    Technology, Rochester, NY 14623, USA\\
    $^6$ Radcliffe Institute for Advanced Study, 10 Garden Street, Cambridge, MA 02138, USA\\
    $^7$ Department of Physics and Astronomy, Michigan State University,
    East Lansing, MI 48824, USA\\
    $^8$ Department of Physics, Durham University, Durham DH1 3LE\\
    $^9$ Department of Physics, University of Texas at Dallas, 800 W Campbell Rd, Richardson, TX 75080, USA\\
    $^{10}$ Department of Physics, Rochester Institute of Technology,
    Rochester, NY 14623, USA\\
  }
}

\maketitle

\begin{abstract}
We present a new $400\ks$ \textit{Chandra} X-ray observation of the
merging galaxy cluster Abell 2146.  This deep observation reveals
detailed structure associated with the major merger event including
the Mach $M=2.3\pm0.2$ bow shock ahead of the dense, ram pressure
stripped subcluster core and the first known example of an upstream
shock in the ICM ($M=1.6\pm0.1$).  By measuring the electron
temperature profile behind each shock front, we determine the
timescale for the electron population to thermally equilibrate with
the shock-heated ions.  We find that the temperature profile behind
the bow shock is consistent with the timescale for Coulomb collisional
equilibration and the postshock temperature is lower than expected for
instant shock-heating of the electrons.  Although like the Bullet
cluster the electron temperatures behind the upstream shock front are
hotter than expected, favouring the instant heating model, the
uncertainty on the temperature values is greater here and there is
significant substructure complicating the interpretation. We also
measured the width of each shock front and the contact discontinuity
on the leading edge of the subcluster core to investigate the
suppression of transport processes in the ICM.  The upstream shock is
$\sim440\kpc$ in length but appears remarkably narrow over this
distance with a best-fit width of only $6^{+5}_{-3}\kpc$ compared with
the mean free path of $23\pm5\kpc$.  The leading edge of the
subcluster core is also narrow with an upper limit on the width of
only $2\kpc$ separating the cool, multiphase gas at $0.5-2\keV$ from
the shock-heated surrounding ICM at $\sim6\keV$.  The strong
suppression of diffusion and conduction across this edge suggests a
magnetic draping layer may have formed around the subcluster core.
The deep \textit{Chandra} observation has also revealed a cool, dense
plume of material extending $\sim170\kpc$ perpendicular to the merger
axis, which is likely to be the disrupted remnant of the primary
cluster core.  This asymmetry in the cluster morphology indicates the
merger has a non-zero impact parameter.  We suggest that this also
explains why the SW edge of the subcluster core is narrow and stable
over $\sim150\kpc$ in length but the NE edge is broad and being
stripped of material.
\end{abstract}

\begin{keywords}
  X-rays: galaxies: clusters --- galaxies: clusters: Abell 2146 --- intergalactic medium
\end{keywords}

\section{Introduction}
\label{sec:intro}

Galaxy clusters are formed through mergers of smaller subclusters and
groups, which collide at velocities of several thousand $\km\s^{-1}$.
The total kinetic energy of these mergers can reach $10^{64}\erg$, a
significant fraction of which is dissipated by large-scale shocks and
turbulence over the merger lifetime (for a review see
\citealt{Markevitch07}).  Shocks and turbulence generated by the
merger are also expected to amplify magnetic fields in the cluster and
accelerate relativistic particles.  These non-thermal phenomena have
been revealed through the detection of Mpc-scale synchrotron radio
halos (for recent reviews see \citealt{Feretti08};
\citealt{Ferrari08}) and inverse-Compton hard X-ray emission
(\citealt{Fusco-Femiano05}; \citealt{Rephaeli99} but see also
\citealt{Wik09}).  The combination of X-ray and gravitational lensing
studies of merging clusters has also produced compelling evidence for
dark matter (eg. \citealt{Clowe04,Clowe06}; \citealt{Bradac06}) and
constraints on the dark matter self-interaction cross section
(eg. \citealt{Randall08}).


\textit{Chandra}'s subarcsecond angular resolution revealed sharp
X-ray surface brightness edges in merging systems.  These edges
correspond to cold fronts or contact discontinuities between regions
of gas with different entropies (\citealt{Markevitch00};
\citealt{Vikhlinin01}; \citealt{Markevitch07}), and shock fronts
driven by infalling subclusters.  However, whilst cold fronts are also
found in relaxed clusters and are common in cluster cores
(eg. \citealt{Owers09}), there are only a few confirmed detections of
shock fronts with a sharp density edge and an unambiguous temperature
jump (the Bullet cluster, \citealt{Markevitch02}; Abell 520,
\citealt{Markevitch05}; two in Abell 2146, \citealt{Russell10}; Abell
754, \citealt{Macario11} and Abell 2744, \citealt{Owers11}).  These
surface brightness edges are key observational tools for studying
merging systems and provide currently the only method for measuring
the bulk velocities of the gas in the plane of the sky and determining
the velocity and kinematics of the merger.

Detailed observations of shock fronts and cold fronts have been used
to probe the relatively unknown transport processes in the
intracluster medium (ICM).  \citet{Markevitch06} used a deep
observation of the bow shock in the Bullet cluster to produce the
first measurement of the electron-ion thermal equilibration timescale
in the ICM and determine that the timescale is likely to be shorter
than the Coulomb collisional timescale (eg. \citealt{Fox97}; \citealt{Markevitch06};
\citealt{Markevitch07}).  This exciting result suggests a heating
process that operates faster than Coulomb collisions could be
operating in the magnetized ICM (eg. \citealt{Schekochihin05};
\citealt{Schekochihin08}).  Observations of the sharp temperature and
density jumps at cold fronts suggest that thermal conduction and
diffusion are strongly suppressed across these edges
(eg. \citealt{Ettori00}).  A detailed study of the cold front in Abell
3667 found that the width of the density jump is smaller than the
Coulomb mean free path in the ICM (\citealt{Vikhlinin01}).  This edge
also appears sharp and stable within a large sector of $\pm30^{\circ}$
around the symmetry axis suggesting that hydrodynamic instabilities
are also suppressed.  The flow of the ambient ICM around the dense
subcluster core will stretch initially tangled magnetic field lines to
form a draping layer with a magnetic field parallel to the front
(\citealt{VikhlininBfield01}; \citealt{Vikhlinin02}; \citealt{Asai05};
\citealt{Lyutikov06}).  This magnetic draping layer could provide a
stabilising mechanism and will suppress transport processes across the
edge.  The long, straight edges of the bullet subcluster in the Bullet
cluster also suggest a strong suppression of turbulence by such a
stabilising magnetic layer (\citealt{Markevitch06}).


The galaxy cluster Abell 2146 ($z=0.234$; \citealt{Struble99};
\citealt{Bohringer00}) is a spectacular merging system with two large
Mach $M\sim2$ shock fronts (\citealt{Russell10}).  The X-ray
morphology suggests a recent collision, $0.1-0.3\Gyr$ ago, where a
subcluster containing a dense cool core has passed through the centre
of a primary cluster.  The subcluster is driving a bow shock through
the ICM and is trailed by a tail of ram pressure stripped material.
An upstream shock is also observed to be propagating in the opposite
direction through the outskirts of the primary cluster.
\citet{Canning11} used the line of sight velocity difference between
the subcluster and primary cluster brightest cluster galaxies (BCGs)
with the shock velocities to estimate that the merger axis is inclined
at only $\sim17^{\circ}$ to the plane of the sky.  This conclusion is
also supported by the sharp surface brightness edges of the bow and
upstream shock fronts, which would be smeared by projection for a
larger angle to the line of sight.  Abell 2146 appears to have
undergone a simple merger between two smaller clusters, viewed close
to side on, and therefore has a remarkably similar structure to the
Bullet cluster (\citealt{Markevitch02}; \citealt{Markevitch06}).

In this paper, we present results from a deep $400\ks$
\textit{Chandra} observation of Abell 2146 studying the transport
processes in the ICM.  In section \ref{sec:chandra}, we discuss the
\textit{Chandra} data reduction, analyse the X-ray morphology and
present maps of the ICM temperature, density and metallicity. In
section \ref{sec:shockfronts}, we analyse the shock fronts in detail and
compare the postshock electron temperature profiles behind the bow and
upstream shock fronts with models for instant and collisional
electron-ion equilibration.  In section \ref{sec:core}, we study the
ram pressure stripping of the subcluster core and study the
suppression of diffusion and conduction across the leading edge of
the subcluster core.  We assume $H_0=70\kmpspMpc$, $\Omega_m=0.3$ and
$\Omega_\Lambda=0.7$, translating to a scale of $3.7\kpc$ per arcsec
at the redshift $z=0.234$ of Abell 2146.  All errors are $1\sigma$
unless otherwise noted.

\section{\textit{Chandra} data analysis}
\label{sec:chandra}

\begin{figure*}
\begin{minipage}{\textwidth}
\centering
\includegraphics[width=0.98\columnwidth]{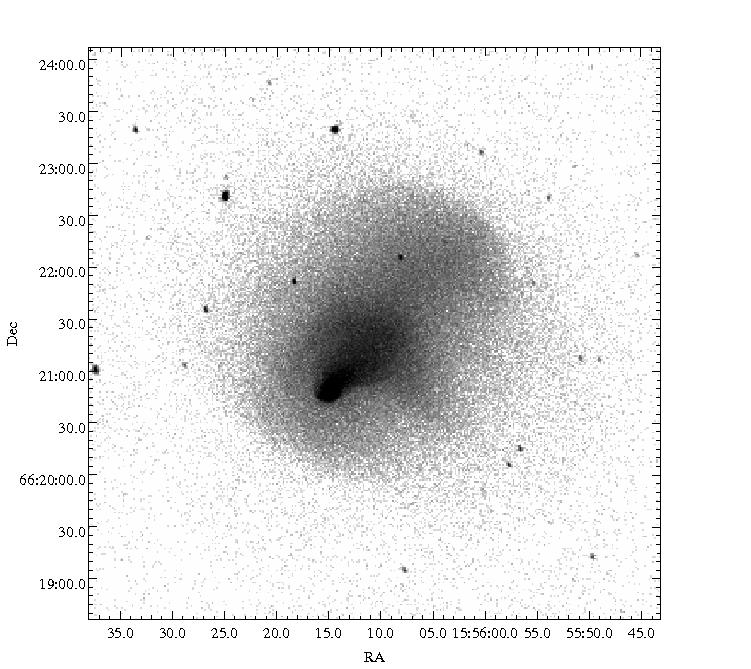}
\caption{Raw counts image of Abell 2146 in the 0.3--7.0$\keV$ energy
  band.  The image has been binned by a factor of 2.}
\label{fig:rawcounts}
\end{minipage}
\end{figure*}

\begin{table}
\caption{Details of the \textit{Chandra} observations analysed in this paper.}
\begin{center}
\begin{tabular}{l c c c c}
\hline
Date & Obs. ID & Aimpoint & Exposure & Cleaned \\
 & & & (ks) & (ks) \\
\hline
2009 April 29 & 10888 & S3 & 6.4 & 6.4 \\
2009 April 30 & 10464 & S3 & 35.7 & 35.7 \\
2010 August 10 & 13020 & I0 & 41.5 & 41.5 \\
2010 August 12 & 13021 & I0 & 48.4 & 48.4 \\
2010 August 19 & 13023 & I1 & 27.7 & 27.7 \\
2010 August 20 & 12247 & I1 & 65.2 & 65.2 \\
2010 September 8 & 12245 & I2 & 48.3 & 48.3 \\
2010 September 10 & 13120 & I2 & 49.4 & 49.4 \\
2010 October 4 & 12246 & I3 & 47.4 & 47.2 \\
2010 October 10 & 13138 & I3 & 49.4 & 48.4 \\
\hline
\end{tabular}
\end{center}
\label{tab:obs}
\end{table}

\subsection{Data reduction}
\label{sec:chandradata}

Abell 2146 was observed with the \textit{Chandra} ACIS-I detector for
a total of $377\ks$ split into eight separate observations between
August and October, 2010 (Table \ref{tab:obs}).  These new
observations were analysed together with the archival ACIS-S observations taken
in April, 2009 (\citealt{Russell10}, Table \ref{tab:obs}).  All
datasets were reprocessed with CIAO 4.3 and CALDB 4.4.0 provided by
the \textit{Chandra} X-ray Center (CXC).  The level 1 event files were
reprocessed to apply the latest gain and charge transfer inefficiency
correction and then filtered to remove photons detected with bad
grades.  The improved background screening provided by VFAINT mode was
also applied.  Background light curves were extracted from the level 2
event files of neighbouring chips for observations on ACIS-I and from
ACIS-S1 for observations on ACIS-S3.  The background light curves were
filtered using the \textsc{lc\_clean} script\footnote{See
  http://cxc.harvard.edu/contrib/maxim/acisbg/} provided by
M. Markevitch to identify periods affected by flares.  There were no
major flares in any of the observations of Abell 2146 producing a
final cleaned exposure of $418\ks$.


The cleaned events files were then reprojected to match the position
of the obs. ID 12245 observation.  Fig. \ref{fig:rawcounts} shows the total
combined image produced by summing images in the $0.3-7.0\keV$ energy
band extracted from each individual reprojected dataset.  This summed
image was then corrected for exposure variation by dividing by the
summed $1.5\keV$ monochromatic exposure maps created for each dataset.
Point sources were identified using the CIAO algorithm
\textsc{wavdetect}, visually confirmed and excluded from the analysis
using elliptical apertures with radii set to five times the measured
PSF width (\citealt{Freeman02}).  

Standard blank-sky backgrounds were extracted for each chip in each
observation, processed identically to the events file and reprojected
to the corresponding sky position.  Each blank-sky background was
normalized to match the count rate in the $9.5-12\keV$ energy band in
the observed dataset.  This correction was less than 10\% for each
dataset.  The normalized blank-sky background events files for each
chip in an observation were then split to ensure each had the same
ratio of exposure time to the observed exposure time.  These were then
summed together to produce background events files which covered all
chips in each observation.  By matching to the hard X-ray background
count rate, we may over or underestimate the soft component of the
X-ray background.  The normalized blank sky background datasets were
tested by comparison with observed background spectra extracted from
source-free regions of the chips.  We found that the normalized blank
sky backgrounds were a close match to the observed background over the
whole energy band.  Total blank-sky background images combining all of
the observations were generated in a similar way to the total images
as detailed above.

\subsection{Imaging analysis}

\begin{figure*}
\begin{minipage}{\textwidth}
\centering
\raisebox{0.07\height}{\includegraphics[width=0.45\columnwidth]{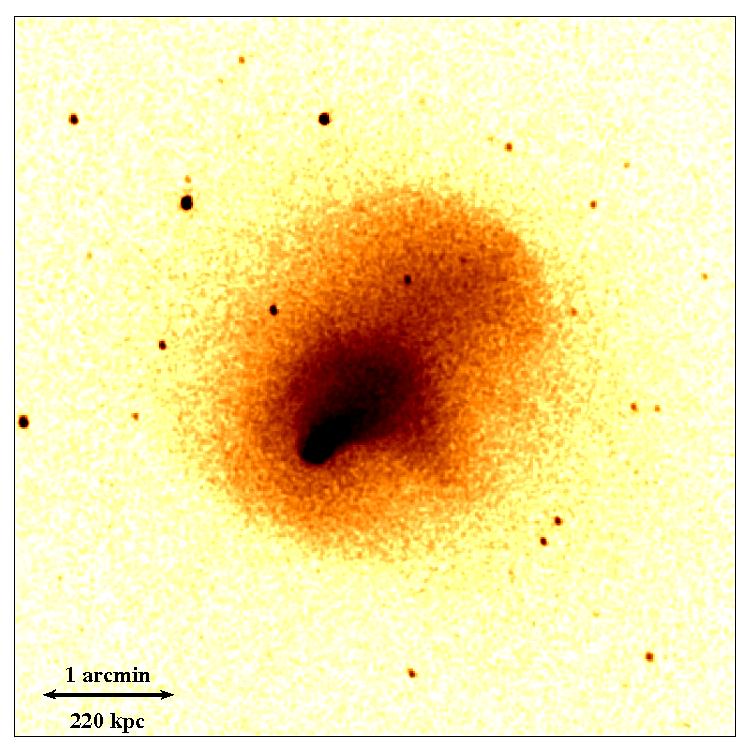}}
\includegraphics[width=0.45\columnwidth]{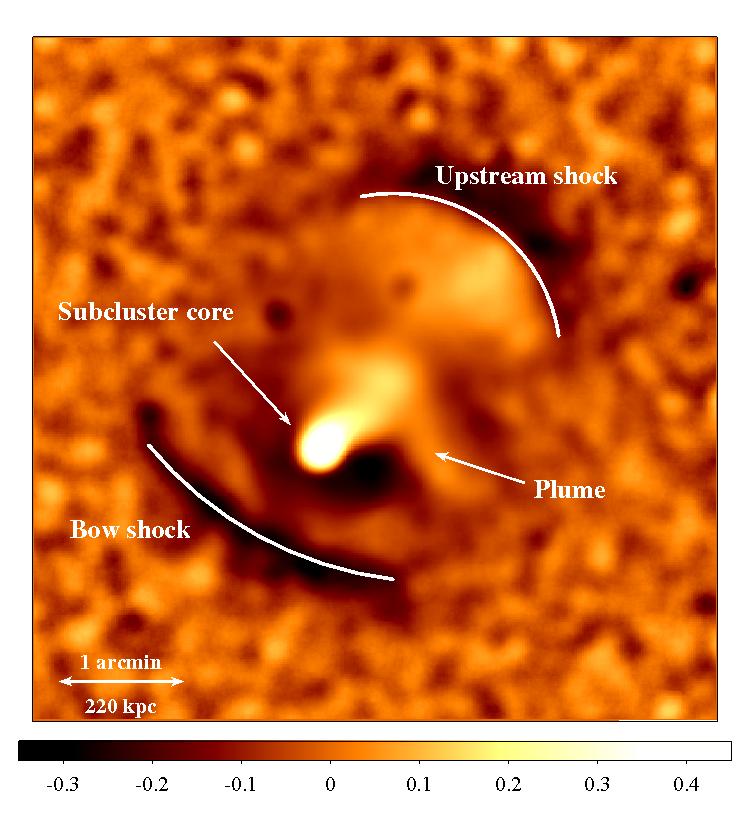}
\includegraphics[width=0.45\columnwidth]{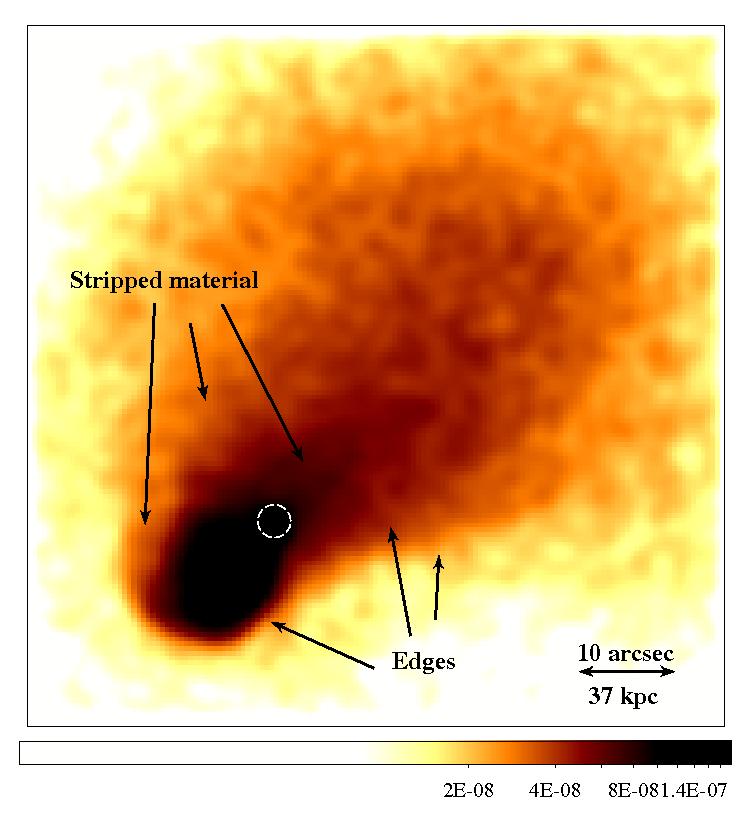}
\raisebox{0.07\height}{\includegraphics[width=0.45\columnwidth]{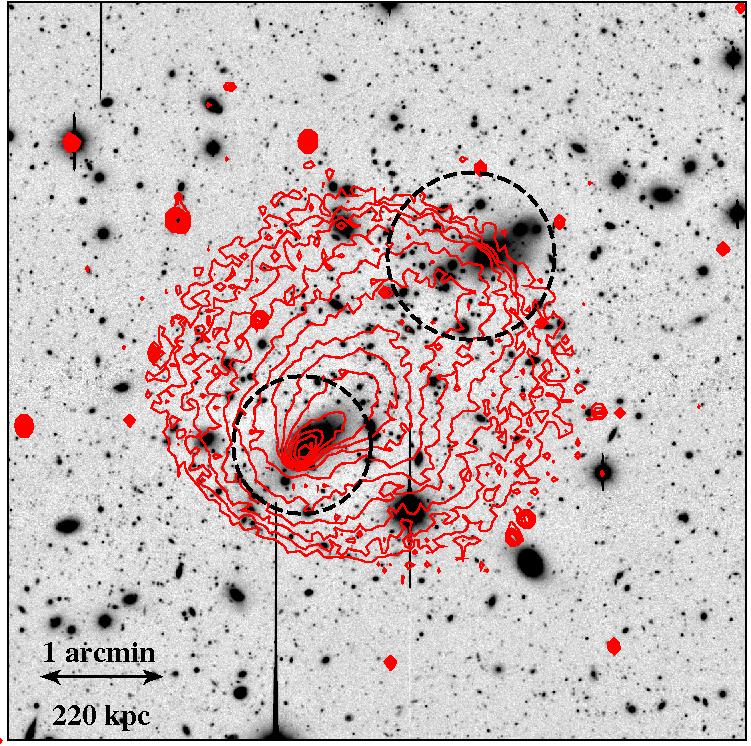}}
\caption{Upper left: Exposure-corrected image in the 0.3--7.0$\keV$
  energy band smoothed with a 2D Gaussian $\sigma=1.5\asec$ (North is
  up and East is to the left).  Upper right: Unsharp-masked image
  created by subtracting images smoothed by 2D Gaussians with
  $\sigma=5$ and $20\asec$ and dividing by the sum of the two images.
  Point sources were removed before unsharp-masking.  Note that a
  point source has been removed at the NE end of the bow shock.  Lower
  left: same as upper left but showing the subcluster core (units \expmapcorr).  The AGN
  nucleus is marked by the white dashed circle.  Lower right: Subaru
  R-band image of the galaxy distribution in Abell 2146 with the X-ray
  gas contours overlaid.  Concentrations of galaxies are marked with dashed circles.}
\label{fig:mainimages}
\end{minipage}
\end{figure*}

Fig. \ref{fig:mainimages} (upper left) shows an exposure-corrected
image of the cluster X-ray emission produced by combining all of the
individual \textit{Chandra} observations.  The cluster gas is extended
along the merger axis, NW to SE, and the bright, dense core of the
subcluster is offset from the centre and being stripped of its
material by ram pressure in the collision.  The X-ray morphology
suggests a major merger where the subcluster has recently passed
through the centre of the primary cluster.  There is no obvious
surface brightness peak associated with a primary cluster core.  The
primary cluster may not have originally had a dense core or it could
have been significantly disrupted in the collision with the subcluster
core.  The subcluster is observed soon after core passage when it has
emerged from the primary core and is travelling towards the SE.
Fig. \ref{fig:mainimages} (lower right) shows a Subaru R-band image of
the galaxy distribution (King et al. in prep.).  The distribution of
red sequence galaxies appears to separate into two groups,
corresponding to the locations of the subcluster and the primary
cluster, which supports the interpretation of a major merger
(\citealt{Canning11}).

The two shock fronts, reported in \citet{Russell10}, are clearly
visible in the unsharp-masked image (Fig. \ref{fig:mainimages}, upper
right) as surface brightness edges to the SE and NW.  The SE edge
corresponds to the bow shock, which has formed ahead of the subcluster
core, and can now be traced over $\sim500\kpc$ in length.  The NW
surface brightness edge corresponds to the upstream shock, which has
formed in the wake of the subcluster's passage through the primary
cluster core and is travelling in the opposite direction to the bow
shock (\citealt{Russell10}).  The upstream shock is visible over
$\sim440\kpc$ in length and appears to have greater curvature compared
to the bow shock.  Note that in Fig. \ref{fig:mainimages} (upper
right) a point source has been removed at the NE end of the bow shock,
which has the effect of increasing the apparent curvature of the shock
front.  There is no obvious second surface brightness peak
corresponding to the primary cluster core and this may have been
completely disrupted in the collision with the subcluster.



The deeper \textit{Chandra} observations have also now revealed the
complex structure of the cool, dense subcluster core and the ram
pressure stripped tail.  Whilst the leading edge of the core is
smooth, narrow and roughly spherical, there is a clear difference
between the NE and SW edges of the tail (Fig. \ref{fig:mainimages},
lower left).  The SW edge appears sharp and narrow over a distance of
$\sim40\asec$.  In comparison, the NE edge is poorly defined, it
appears broader and disrupted, suggesting that the interface with the
ambient medium has become turbulent here and instabilities could be
developing.  There is also an extended plume of emission to the SW
from the subcluster tail, $\sim45\asec$ in length, which is
perpendicular to the merger axis.  These features are discussed
further in section \ref{sec:core}.

\begin{figure}
\centering
\includegraphics[width=0.95\columnwidth]{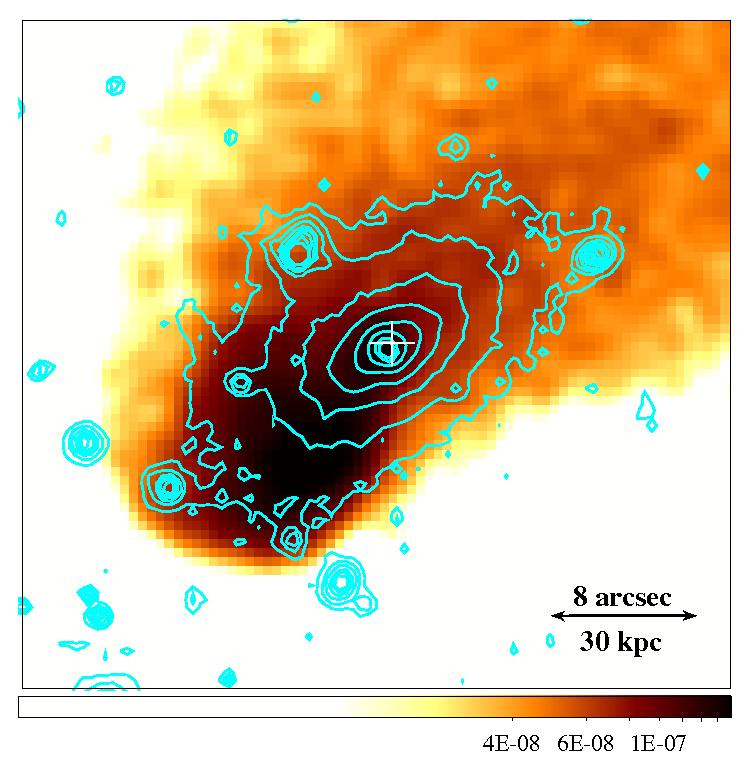}
\caption{Exposure-corrected image of the subcluster core in the
  0.3--7.0$\keV$ energy band smoothed with a 2D Gaussian
  $\sigma=1\asec$ (North is up and East is to the left; units
  \expmapcorr).  The blue contours were produced from the \textit{HST}
  F606W archival image of the galaxies (blue), including the BCG, and
  the white cross shows the location of the VLA $1.4\GHz$ point source
  marking the nucleus (NRAO/VLA Archive Survey).}
\label{fig:AGNandBCG}
\end{figure}

Fig. \ref{fig:AGNandBCG} shows the subcluster core in detail and the
location of the BCG immediately behind the X-ray peak.  The deep
\textit{Chandra} dataset confirms the detection of a hard X-ray
($4-7\keV$) point source, at the $3\sigma$ level, coincident with a
radio point source detected in VLA $1.4\GHz$ archival data (NRAO/VLA
Archive Survey) and observations with the AMI Large Array at $16\GHz$
(\citealt{Rodriguez11}).  Also, \textit{Spitzer} observations of the
BCG by \citet{Quillen08} suggest that there is a strong contribution
to the IR emission from an AGN.  This radio and X-ray point source
likely corresponds to an AGN at the centre of the BCG.  However it was
difficult to determine the flux as the nucleus is superimposed on a
bright, dense filament of gas, which is detected in soft X-rays with
\textit{Chandra} but also in H$\alpha$ and [N~\textsc{ii}]
(\citealt{Canning11}).  We extracted the AGN source counts in a region
of $2\asec$ radius and subtracted the cluster emission using a
neighbouring region to the SE from $2.5-5$ radius, which lay on the
dense gas filament.  For the energy band $2-7\keV$, the point source
was detected as $210\pm30$\,counts above the background cluster
emission.  Assuming a powerlaw model with photon index $\Gamma=2$ and
Galactic absorption of $n_{\mathrm{H}}=0.03\times10^{22}\pcmsq$
(\citealt{Kalberla05}), we estimated the point source luminosity in
the energy range $2-10\keV$ to be
$L_{X}=1.5\pm0.2\times10^{42}\ergps$.

\subsection{Spatially resolved spectroscopy}
\label{sec:spectro}

\subsubsection{Contour binning maps}
\label{sec:speccontour}

\begin{figure*}
\begin{minipage}{\textwidth}
\centering
\includegraphics[width=0.45\columnwidth]{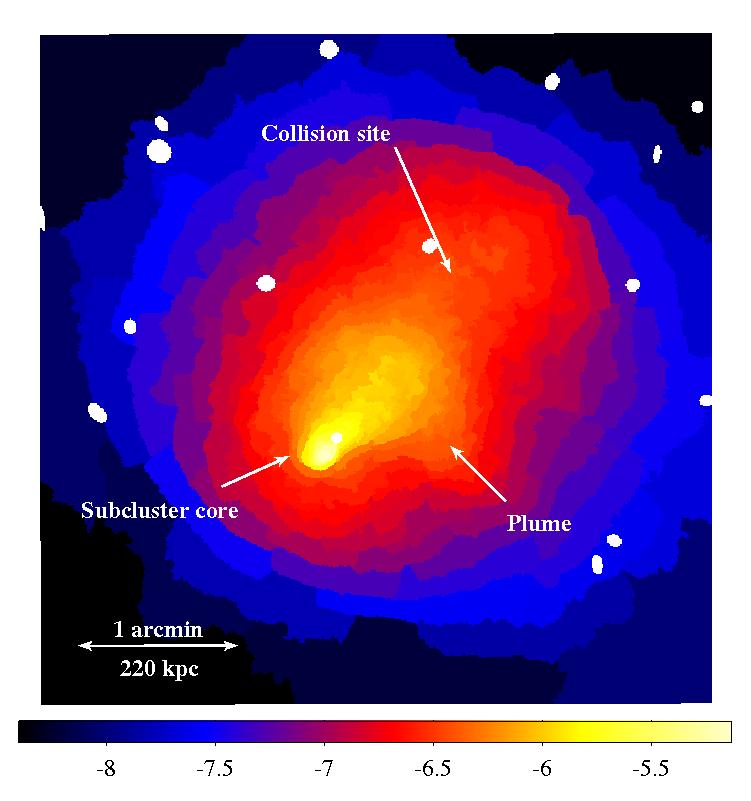}
\includegraphics[width=0.45\columnwidth]{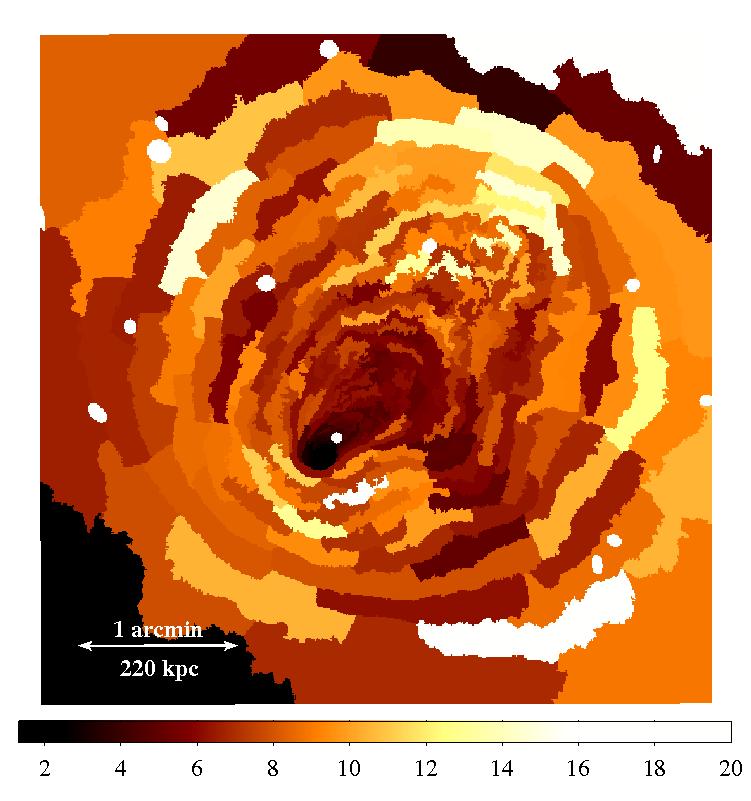}
\includegraphics[width=0.45\columnwidth]{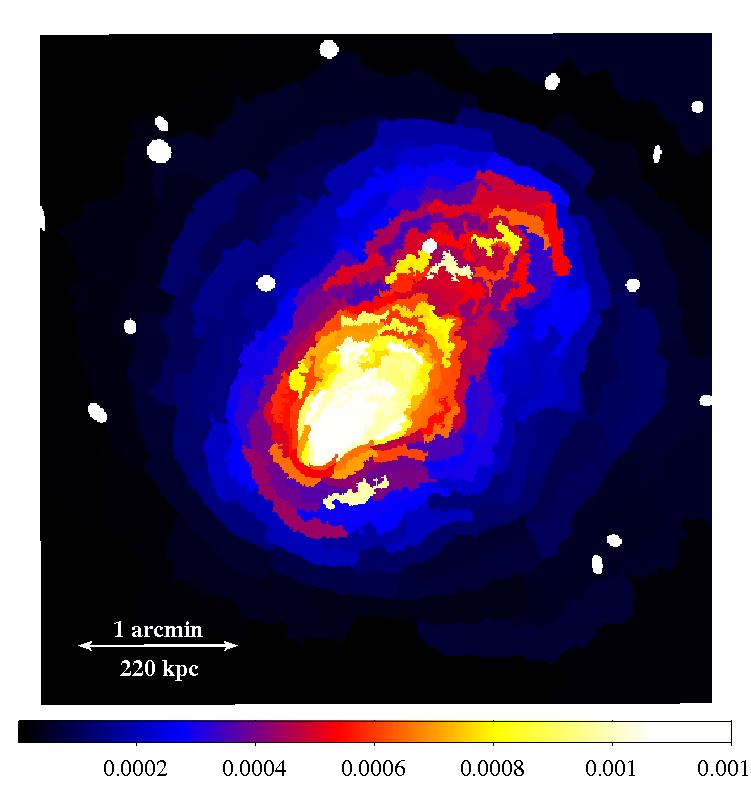}
\includegraphics[width=0.45\columnwidth]{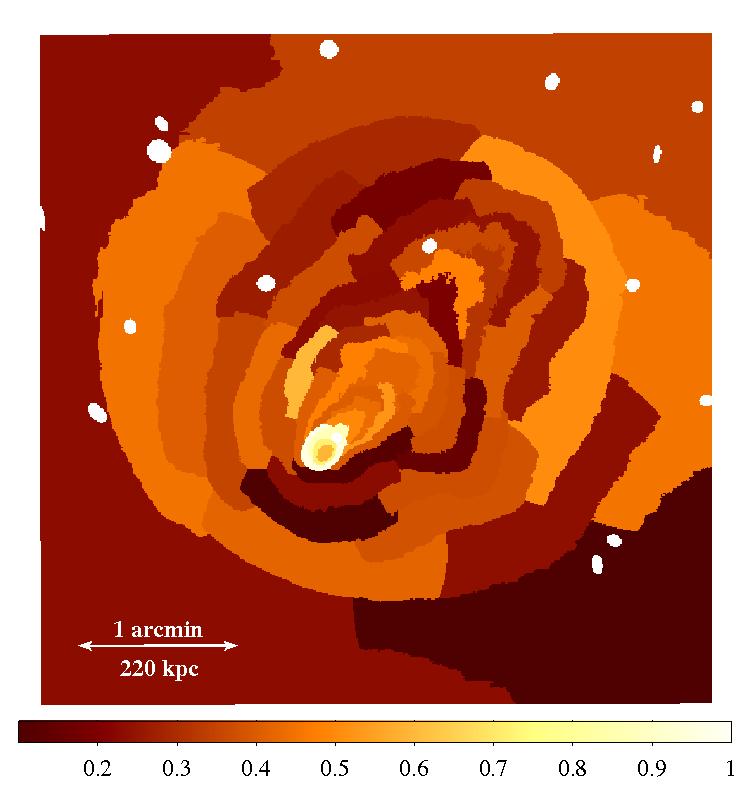}
\caption{Upper left: projected emission measure per unit area map
  (units are log$_{10}\empasecsq$).  The emission measure is the
  \textsc{xspec} normalization of the \textsc{mekal} spectrum
  $K=EI/(4\times10^{14}\pi D_A^2(1+z)^2)$, where EI is the emission
  integral $EI=\int n_en_H\mathrm{d}V$.  The approximate position of
  the core collision site is labelled.  Upper right: projected
  temperature map (keV).  Note that the white circle behind the cool
  core is the excluded central AGN.  Lower left: projected
  pseudo-pressure map ($\pseudoP$).  Lower right: projected
  metallicity map ($\Zsun$) generated using larger spectral bins with
  $\sim 5000$ counts.  The excluded point sources are visible as small
  white circles.  North is up and East is to the left.}
\label{fig:maps}
\end{minipage}
\end{figure*}

We used spatially resolved spectroscopy techniques to produce detailed
maps of the projected gas properties (Fig. \ref{fig:maps}).  The
central $\sim4\times4\amin$ was divided into regions using the Contour
Binning algorithm (\citealt{Sanders06}), which traces the surface
brightness variations to generate the spatial bins.  For the
temperature and normalization maps, regions with a signal-to-noise
ratio of 32 ($\sim1000$ counts) were chosen.  The length of the
regions was restricted to be at most two and a half times their width.
For each \textit{Chandra} observation, we extracted a spectrum from
each of the regions, subtracted the background spectrum and generated
appropriate responses and ancillary responses.  The spectra were
grouped to contain a minimum of 20 counts per spectral channel and
restricted to the energy range $0.5-7\keV$.  We summed together
spectra and background spectra for a particular region for
observations on the same chip, as the roll angles are also comparable
in each case.  The response files were also averaged together,
weighting by the fraction of the total counts in each observation.
Each total spectrum was fitted in \textsc{xspec} 12
(\citealt{Arnaud96}) with an absorbed \textsc{mekal} model.  The
absorption was fixed to the Galactic value $n_{\mathrm
  H}=3.0\times10^{20}\pcmsq$ (\citealt{Kalberla05}) and the redshift
was fixed to 0.234.  The C-statistic was minimised in the spectral
fitting (\citealt{Cash79}).  The errors are approximately $\sim6\%$ in emission measure
and $\sim15\%$ in temperature.  However, the error in temperature
drops to less than 10\% in the subcluster core, where the temperature
falls below $2\keV$ and the Fe L line emission improves temperature
diagnostics.  In addition, temperatures over $10\keV$ are poorly
constrained by the energy range of \textit{Chandra} and the errors
increase to $\sim30\%$ in these bins.


Fig. \ref{fig:maps} (upper left) shows the distribution of the
projected emission measure, which traces the square of the gas density
in the cluster.  The emission measure peaks on the subcluster core and
there is clearly a sharp density drop at the core's leading edge to
the SE.  The emission measure declines more gradually through the ram
pressure stripped tail of gas to the NW of the core and there is a
plume of emission extending to the SW perpendicular to the merger axis
(Fig. \ref{fig:mainimages}, upper left).  There is no obvious second
peak in the emission measure map corresponding to the primary cluster
core, although there are clumps of more dense material in the
approximate position of the main collision site.  

The temperature map (Fig. \ref{fig:maps}, upper right) shows that the
dense subcluster core contains the coolest gas in the cluster, down to
$1.45\pm0.07\keV$.  The steep density drop of the SE edge of the
cluster core corresponds to a sharp increase in the gas temperature
from $\sim2\keV$ to $\sim6\keV$.  This is a cold front or contact
discontinuity created as the subcluster's dense cool core travels
through the hotter, diffuse outskirts of the primary cluster
(\citealt{Markevitch00}; \citealt{Vikhlinin01}).  The temperature
increases more steadily through the ram pressure stripped tail up to
$\sim6-7\keV$.  The plume extending to the SW is cooler than its
surroundings with a temperature of $5-6\keV$.  There does not appear
to be a similar cool, dense plume structure on the other side of the
subcluster tail to the NE.

The hottest gas in the cluster, at temperatures from $12-15\keV$, is
located at the core collision site to the NW.  The NW edge of this
high temperature region is the upstream shock and corresponds to a
drop in the gas density shown by the emission measure map.  The bow
shock is also visible as a peak in the temperature map to the SE but
this is not as clearly shown by the selected binning.  

The projected `pressure' map (Fig. \ref{fig:maps}, lower left) was
produced by multiplying the square root of the emission measure and
the temperature maps.  The `pressure' map reveals the extent of the
shock heating and compression along the merger axis and the sharp
drops in pressure to the NW and SE correspond to the upstream and bow
shock fronts, respectively.  The SW plume appears to be in pressure
equilibrium with its surrounding environment and this is explored
further in section \ref{sec:specradial}.  Although there are regions
of high pressure that could be associated with the core collision
site, it was difficult to determine this location exactly and the
formation of the plume indicates it could instead be closer to the
subcluster tail (section \ref{sec:specradial}).

The signal-to-noise ratio was increased to 70 ($\sim5000$ counts) to
create the metallicity map.  The errors on the metallicity values are
approximately $\pm0.08\Zsun$ but increase to $0.1-0.15\Zsun$ for the
highest temperature shock-heated regions where the ICM is almost completely
ionised and there is minimal line emission.  The metallicity peaks in
the subcluster core, ahead of the BCG, at $\sim0.9\Zsun$ and is
approximately constant elsewhere at $\sim0.4\Zsun$.  The apparent drop
in the metallicity inside the subcluster core is an artifact caused by
the use of a single temperature spectral model where the cluster gas
has multiple temperature components (\citealt{Buote94};
\citealt{Buote00}).  If we fit the spectra from this region with a two
temperature model, the best-fit metallicity value is
$0.88^{+0.10}_{-0.09}\Zsun$, which is consistent with the average for
the subcluster core.  Two temperature model fits are discussed in more
detail in section \ref{sec:core}.  

The metallicity drops sharply across the SE edge of the subcluster
core showing clearly the difference in origin of the gas on either
side of this contact discontinuity.  It is not clear why the postshock
gas ahead of the subcluster core has a lower average metallicity.  By
fitting the spectra from the three low metallicity regions ahead of
the subcluster core together, we find a best-fit value of only
$0.15\pm0.06\Zsun$ compared to the ambient value of $\sim0.4\Zsun$.
Although the gas temperature is high here and the metallicity is more
difficult to constrain, we note that the metallicity for the
shock-heated gas behind the upstream shock has a metallicity
consistent with the average at $0.5\pm0.2\Zsun$.  There was also no
evidence from spectral fitting for a second temperature component in
this region.  We consider the possible impact of non-equilibrium
ionization behind the shock fronts in section \ref{sec:nei}.  The metallicity
also drops rapidly behind the subcluster core from $0.9\pm0.08\keV$ to
$0.5^{+0.08}_{-0.07}\Zsun$ in the region immediately behind the AGN.
There is no evidence for a metallicity gradient in the ram pressure
stripped tail suggesting that the metal enriched material is pulled
off the subcluster core in clumps which don't efficiently mix with the
ambient ICM.

\subsubsection{Radial profiles}
\label{sec:specradial}

\begin{figure}
\centering
\includegraphics[width=0.8\columnwidth]{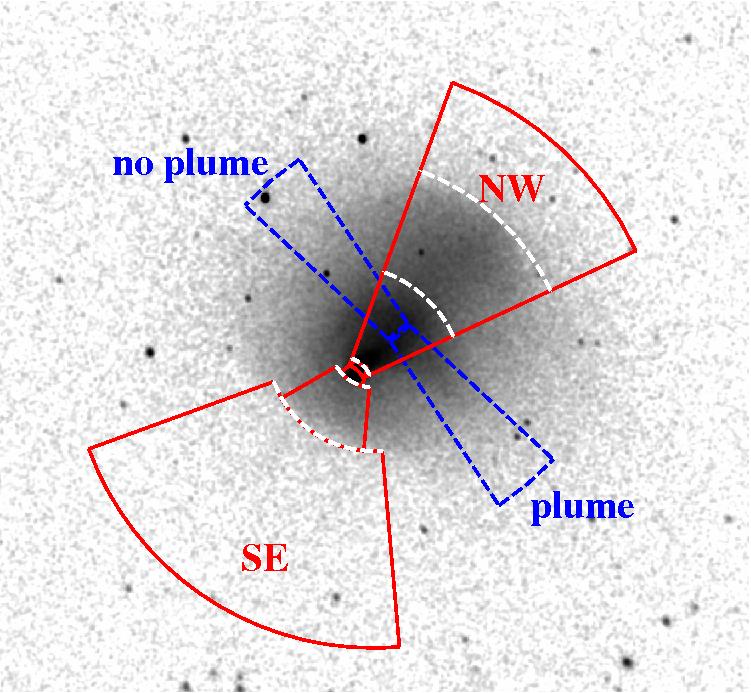}
\caption{Image showing the sectors used to produce the radial profiles
  in Fig. \ref{fig:sectorprofiles} and Fig. \ref{fig:plumeprofiles}
  with the surface brightness edges marked by dashed white lines.}
\label{fig:specsectors}
\end{figure}
\begin{figure*}
\begin{minipage}{\textwidth}
\centering
\includegraphics[width=0.45\columnwidth]{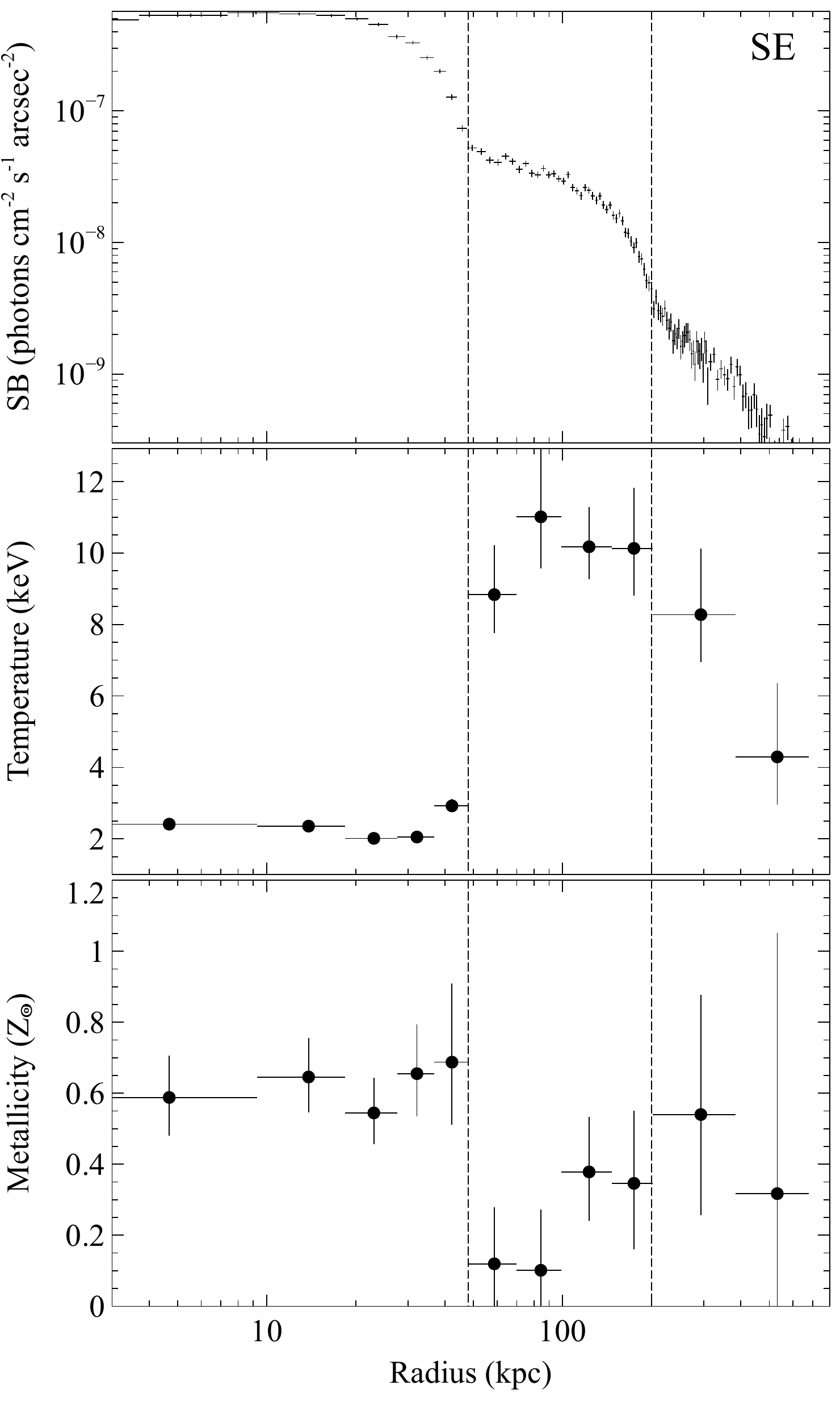}
\includegraphics[width=0.45\columnwidth]{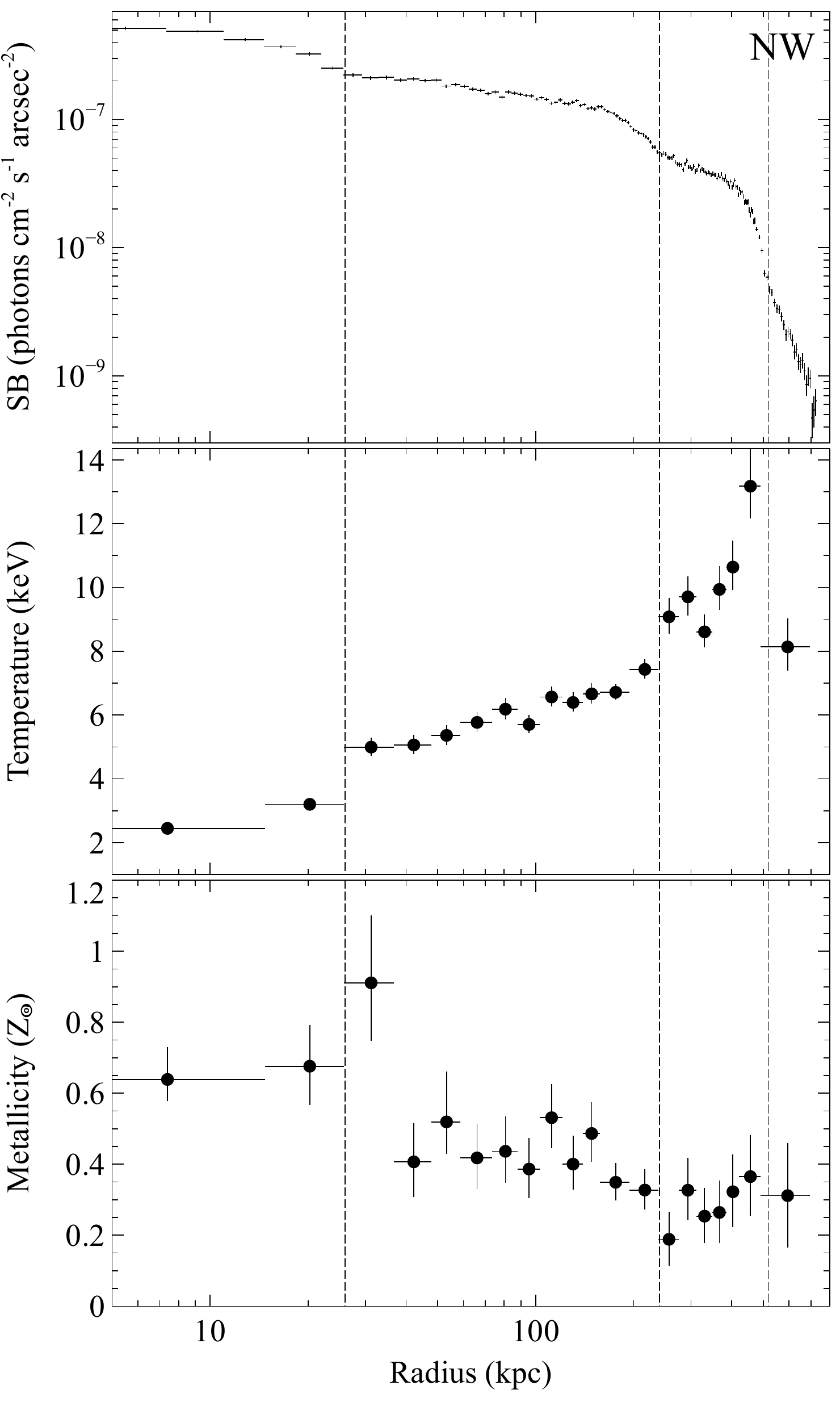}
\caption{Surface brightness, temperature and metallicity profiles for
  the SE (left) and NW (right) sectors shown in
  Fig. \ref{fig:specsectors}.  The surface brightness profiles cover
  the energy range $0.5-7\keV$ and edges, corresponding to cold fronts
  or shocks, are marked by dashed lines.  Note that the sectors were
  not selected to correspond to the shock edges (see
  Fig. \ref{fig:paramprofiles}).}
\label{fig:sectorprofiles}
\end{minipage}
\end{figure*}

Several key sectors of the merging cluster are identified in
Fig. \ref{fig:specsectors} and used to investigate these structures in
greater detail.  We extracted projected surface brightness,
temperature and metallicity profiles in these regions using a minimum
of 50 source counts per radial bin for the surface brightness and 2000
counts per bin for the spectral fitting.  The temperature and
metallicity were determined by fitting a single temperature model to
each extracted spectrum as in section \ref{sec:speccontour}.
Fig. \ref{fig:sectorprofiles} shows the radial profiles for the SE and
NW sectors positioned along the merger axis.  The SE sector
(Fig. \ref{fig:specsectors}) is centred on the subcluster core and
includes the leading edge of the core and the bow shock.  The NW
sector was selected to cover the subcluster tail, the collision site
and the upstream shock.  Note that the centre of each sector was not
the centre of curvature of each shock and so these features appear
smoothed here compared to the later analysis in section
\ref{sec:shocks}.

The surface brightness edges corresponding to the shock fronts and the
contact discontinuity ahead of the subcluster core are clearly visible
in Fig. \ref{fig:sectorprofiles} but we also note two additional edges in the
NW sector immediately behind the subcluster core at $26\kpc$ and at
the end of the subcluster tail at $240\kpc$.  The surface brightness
drop immediately behind the subcluster core is also accompanied by an
increase in the gas temperature from $3.2\pm0.1\keV$ to
$5.0\pm0.3\keV$.  Interestingly, the AGN and centre of the BCG is
located almost exactly at the position of this surface brightness and
temperature edge.  The AGN was excluded from the spectral fits using a
region of radius $2\arcsec$ and therefore did not contribute to the
temperature increase at this location.  It is more likely that this
contact discontinuity has been created by the convergence of the
surrounding ICM behind the dense subcluster core.  The metallicity
immediately behind the core appears consistent with the core value,
although with large errors.  The metallicity then decreases to
$0.4\Zsun$ and is approximately constant through the subcluster tail.
Metal-rich material that is ram pressure stripped from the subcluster
core may build up in the region immediately behind the core where the
flow of the ambient ICM converges.

The surface brightness edge at the end of the subcluster tail (at
$240\kpc$) is broad, covering $\sim80\kpc$ in radius, and has a
relatively shallow surface brightness gradient that does not appear to
be characteristic of a contact discontinuity.  The temperature
gradient increases from $6.7\pm0.2\keV$ to $9.1^{+0.6}_{-0.5}\keV$
across this region.  However, there is likely to be a significant
component of projected emission on this region which will increase the
apparent temperature.  Deprojection routines usually assume spherical
symmetry which is not a good assumption for this region of Abell 2146
where the cluster emission is strongly elongated along the merger axis
and the surface brightness gradient is relatively shallow.  This
surface brightness edge marks the region where $6-7\keV$ material from
the end of the subcluster tail could be mixing with gas from the primary cluster.

\begin{figure*}
\begin{minipage}{\textwidth}
\centering
\includegraphics[width=0.45\columnwidth]{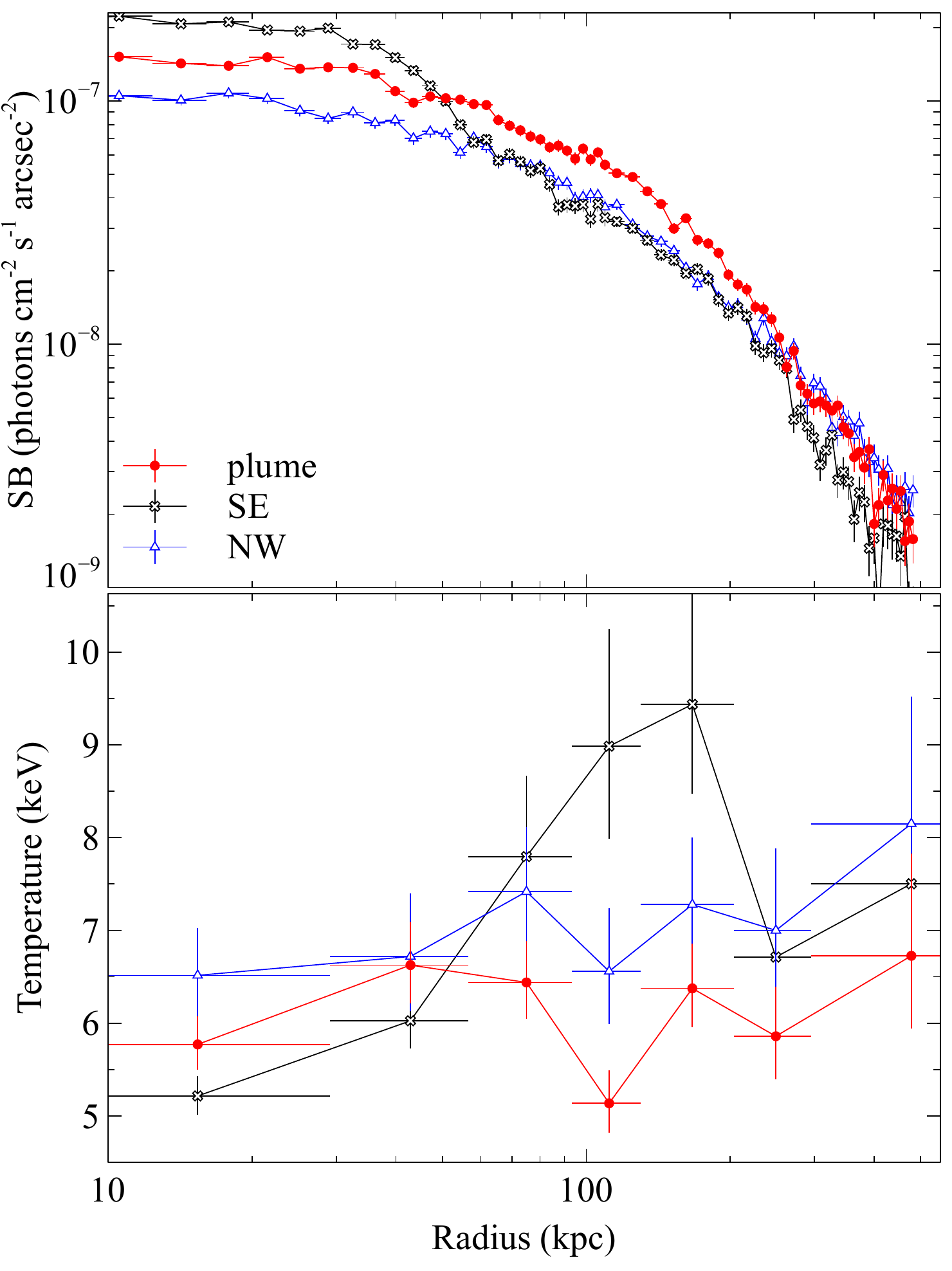}
\includegraphics[width=0.45\columnwidth]{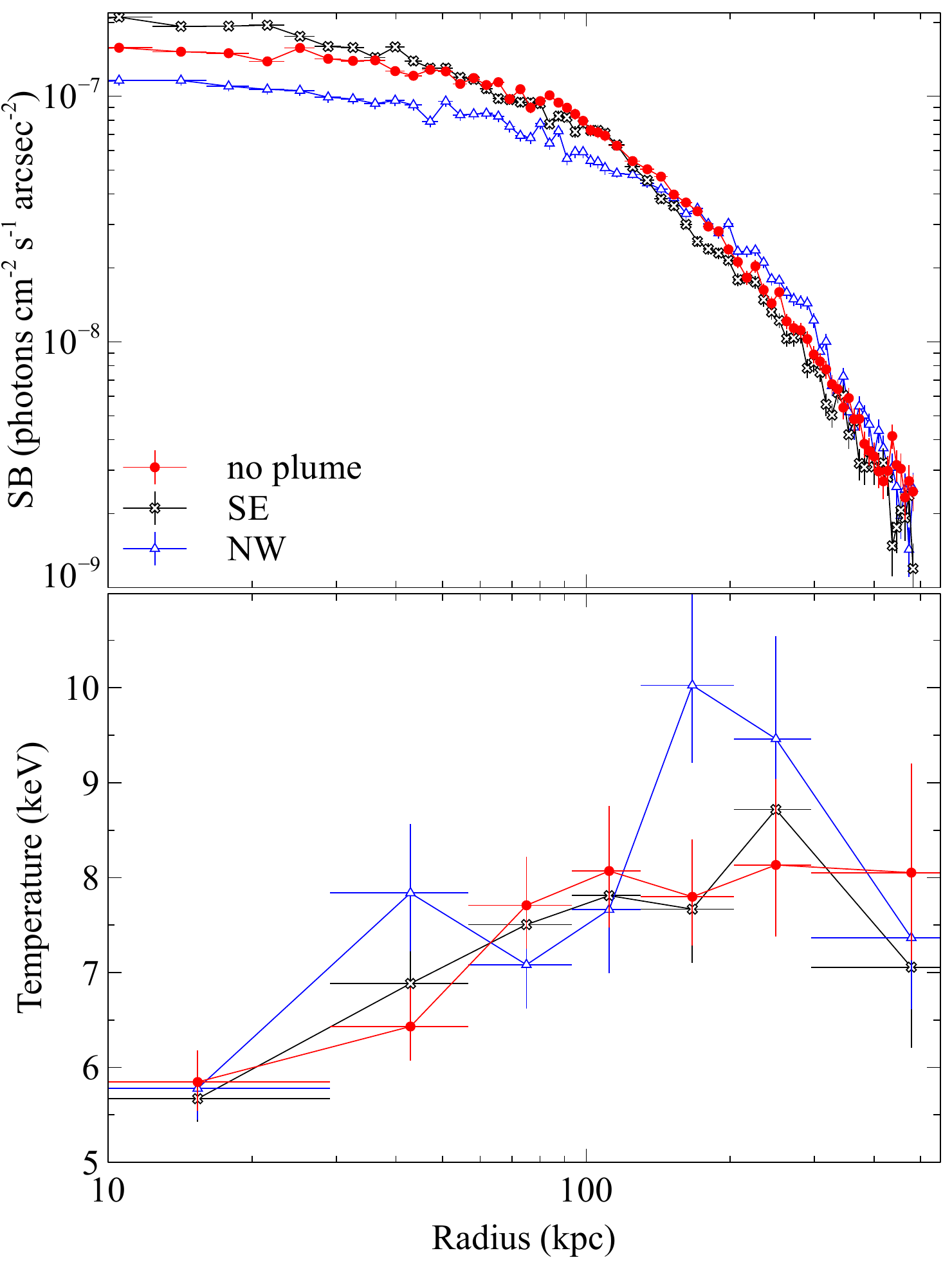}
\caption{Surface brightness and temperature profiles for the plume and
  no-plume sectors shown in Fig. \ref{fig:specsectors}.  The NW and SE
  profiles refer to additional, comparison sectors positioned either
  side of the plume and no-plume sectors.}
\label{fig:plumeprofiles}
\end{minipage}
\end{figure*}

Fig. \ref{fig:plumeprofiles} shows the surface brightness and
temperature of the gas in the SW plume and a comparison no-plume
sector on the other side of the subcluster tail (see
Fig. \ref{fig:specsectors}).  These two sectors were also each
compared with two neighbouring sectors positioned either side to the
NW and SE.  Fig. \ref{fig:plumeprofiles} (left) shows the surface
brightness enhancement of the plume over the surrounding sectors from
$\sim50-250\kpc$.  The electron pressure is possibly higher in the
sector to the SE of the plume as the temperature is higher here but
the errors are large, otherwise the plume is in pressure equilibrium
with the surroundings.  There is no comparable surface brightness
increase in the no-plume sector.  The increase in temperature in the
NW no-plume sector is probably due to the inclusion of a section of
the upstream shock in the region analysed.  The metallicity is
approximately constant at $0.4\Zsun$, similar to the ambient ICM, for
both plume and no-plume sectors.

We suggest that the plume is likely to be the remnant of the primary
cluster core which has been pushed forward and laterally by the impact
of the subcluster core.  There does not appear to be a symmetric
structure extending from the merger axis to the NE but this can be
explained if the two clusters collided with a non-zero impact
parameter (see eg. \citealt{Roettiger98}; \citealt{Ricker01};
\citealt{Poole06}).  Simulations of off-axis mergers indicate that if
the subcluster passed to the North of the primary cluster centre, at a
distance of order the core size, material from the primary
core would be ejected perpendicular to the merger axis in the SW
direction, as observed in Abell 2146.  A strongly off-axis collision,
comparable to the lateral extent of the shock fronts, seems unlikely
however as the two shock fronts are both close to symmetric about the
merger axis and the primary cluster core has been strongly disrupted
by the merger.  Detailed simulations of the merger in Abell 2146 will
be needed to improve these rough limits.



\section{The shock fronts}
\label{sec:shockfronts}
\subsection{Surface brightness profiles}
\label{sec:shocks}

\begin{figure}
\centering
\includegraphics[width=0.8\columnwidth]{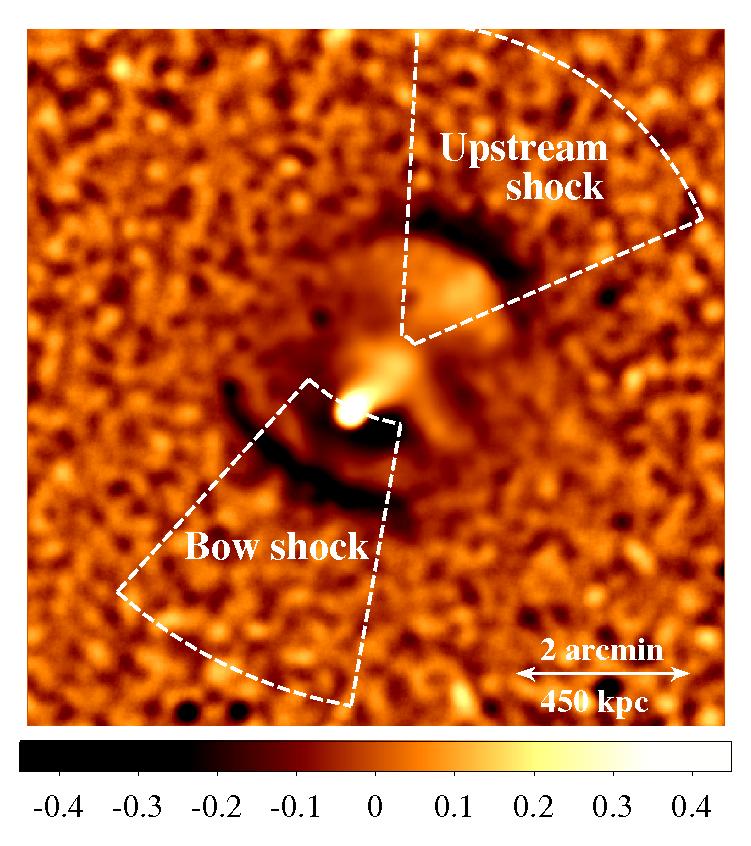}
\caption{Unsharp-masked image (as in Fig. \ref{fig:mainimages}, upper right)
  showing the sectors used to produce surface brightness profiles for
  the shock fronts.  The sectors were centred on the centre of
  curvature for each shock front.}
\label{fig:SBsectors}
\end{figure}

\begin{figure*}
\begin{minipage}{\textwidth}
\centering
\includegraphics[width=0.45\columnwidth]{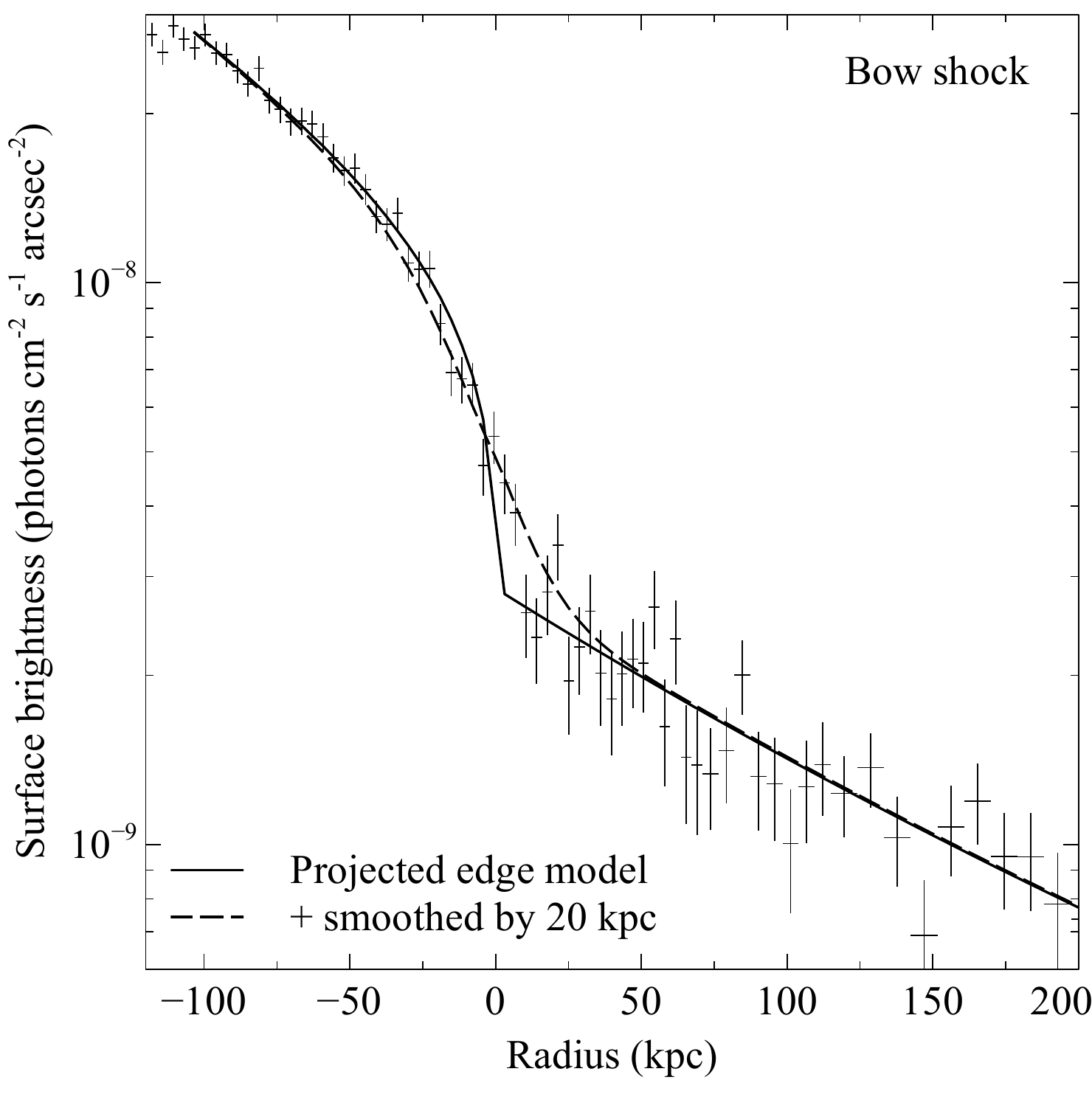}
\includegraphics[width=0.45\columnwidth]{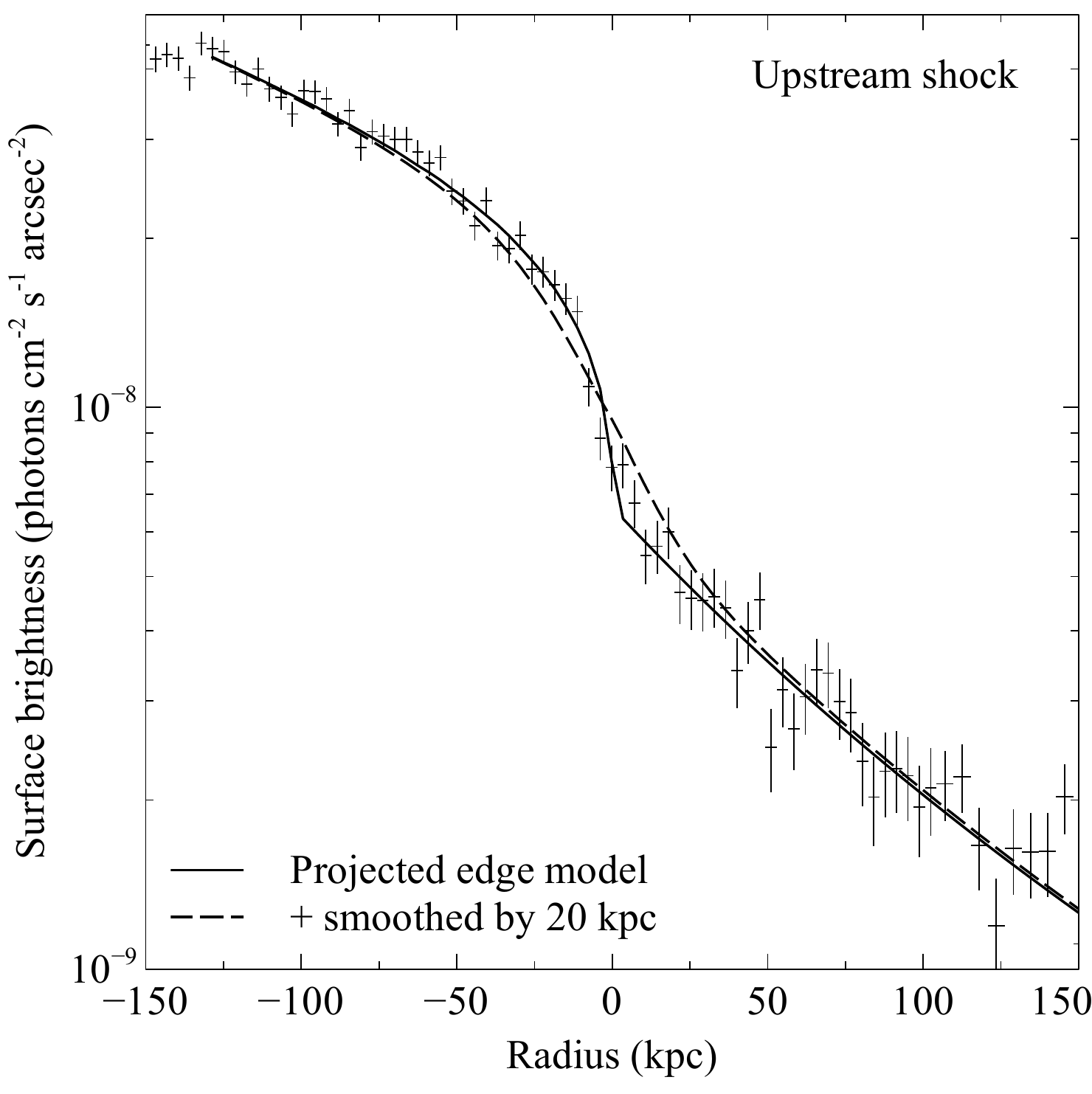}
\caption{Left: Background-subtracted surface brightness profile across
  the bow shock in the energy band 0.5--4.0$\keV$ overlaid with the
  best fit projected density discontinuity model with no smoothing
  (solid lines) and with $20\kpc$ smoothing, corresponding to the mean
  free path $\lambda_{\mathrm{in{\rightarrow}out}}$ (dashed lines).
  Right: same as left but for the upstream shock.}
\label{fig:SBprofiles}
\end{minipage}
\end{figure*}

Fig. \ref{fig:SBsectors} shows the sectors selected for the analysis
of the bow and upstream shock fronts.  The sectors were positioned
according to the centre of curvature of the edge and cover an angular
range where the shock front is well-defined.  These sectors were
divided into radial bins initially of $1\asec$ width but this was
increased to 1.5, 2.5 and $5\asec$ at larger radii to ensure a minimum
of 50 source counts in each bin.  Point sources were excluded from the
analysis and the background was subtracted using the normalized blank
sky background datasets detailed in section \ref{sec:chandradata}.  To
maximise the signal-to-noise ratio, the surface brightness profiles
were restricted to the energy range $0.5-4.0\keV$ and we used only the
\textit{Chandra} observations taken with ACIS-I, which have a lower
background than ACIS-S.  Fig. \ref{fig:SBprofiles} shows the final
background-subtracted X-ray surface brightness profiles for the two
shock fronts.  We fit the surface brightness edge associated with each
shock with a model for a projected spherical density discontinuity,
assuming that the edge is symmetric about the merger axis
(eg. \citealt{Markevitch00,Markevitch02}; \citealt{Owers09}).  The
model radial gas density profile consists of a power law on either
side of an abrupt density jump where the free parameters are the
slopes and normalizations of the power laws and the radial position of
the density jump.  The corresponding emission measure profile was
projected onto the sky and fitted to the observed surface brightness
profile over a radial range selected to exclude unrelated core
structures but extending to large radii for correct projection.

Fig. \ref{fig:SBprofiles} shows that this model provides a reasonable
fit to both the bow and upstream shock fronts.  The amplitude of the
density jump at each shock was calculated from the square root of the
ratio of the power law normalizations (eg. \citealt{Owers09}) with a
small correction for the observed temperature difference across the
edge.  Following \citet{LandauLifshitz59}, we applied the
Rankine-Hugoniot shock jump conditions across each shock front to
calculate the respective Mach numbers, $M=v/c_s$, where $v$ is the
velocity of the preshock gas with respect to the shock surface and
$c_s$ is the velocity of sound in that gas.  The Mach number was
calculated from the density jump,

\begin{equation}
M=\left(\frac{2 \ ^{\rho_2}/_{\rho_1}}{\gamma + 1 - \
    ^{\rho_2}/_{\rho_1}(\gamma -1)}\right)^{1/2},
\label{eq:Mdensity}
\end{equation}

\noindent where $\rho_1$ and $\rho_2$ denote the gas density upstream
and downstream of the shock respectively\footnote{Note that eq. 2 in
  \citet{Russell10}, relating the Mach number to the temperature jump,
  is applicable only for strong shocks.  Given the large error on the
  temperature values in \citet{Russell10} this had a negligable effect
  but as the temperature errors are much smaller for this deeper
  \textit{Chandra} observation a similar assumption is not made
  here.}.  We assume the adiabatic index for a monatomic gas,
$\gamma=5/3$.

At the bow shock, the density drops by a factor
$\rho_2/\rho_1=2.5\pm0.2$ which gives a Mach number $M=2.3\pm0.2$.
For the upstream shock, the density decreases by $\rho_2/\rho_1=1.9\pm0.2$
producing $M=1.6\pm0.1$.  

Even in the raw counts image shown in Fig. \ref{fig:rawcounts} it is
clear that the upstream shock in Abell 2146 has a very narrow edge
separating the pre- and postshock gas.  We estimated the width of the
shocks by smoothing the best-fit density discontinuity model with
Gaussian functions of varying widths, $\sigma$, and fitting the
smoothed models to the surface brightness profile across the shock
front.  For the upstream shock, the best-fit model with zero width has
$\chi^2=98.4$ for 85 degrees of freedom.  This is reduced to
$\chi^2=88.9$ for 84 degrees of freedom for a smoothed model with a
width $\sigma=6^{+5}_{-3}\kpc$ (95\% errors).
Fig. \ref{fig:SBprofiles} shows that the bow shock is broader than the
upstream shock with a best-fit width of $12^{+6}_{-5}\kpc$ (95\%
errors).

The shock widths can be compared with the Coulomb mean free path of
the electrons and protons on both sides of the shock front.  Following
\citet{Vikhlinin01}, we estimate the mean free path for particles
crossing from the postshock gas into the preshock gas,
$\lambda_{\mathrm{in{\rightarrow}out}}$, which is the main source of
diffusion across the edge.  For the bow shock, the mean free path
$\lambda_{\mathrm{in{\rightarrow}out}}=21\pm3\kpc$ and for the
upstream shock it is $23\pm5\kpc$.  If Coulomb diffusion is not
suppressed across the shock edge, we expect the shock to have a width
of several times the mean free path.  The upstream shock is
significantly narrower than the estimated mean free path, which
suggests there is a significant suppression of transport processes
across this edge consistent with a collisionless shock.

The bow shock appears broader however the measured width of the shock
fronts can be affected by any deformations in the front shape, which
smear the edge in projection.  It is therefore difficult to determine
if the bow shock front is intrinsically wider.  The estimated width of
the Bullet cluster bow shock is $\sim35\kpc$, although
\citet{Markevitch06} find this is only marginally preferred over a
zero width shock.  It is however surprising that the upstream shock
front should be the narrowest when it is propagating through the
remaining infall from the subcluster's passage.

\begin{figure*}
\begin{minipage}{\textwidth}
\centering
\includegraphics[width=0.45\columnwidth]{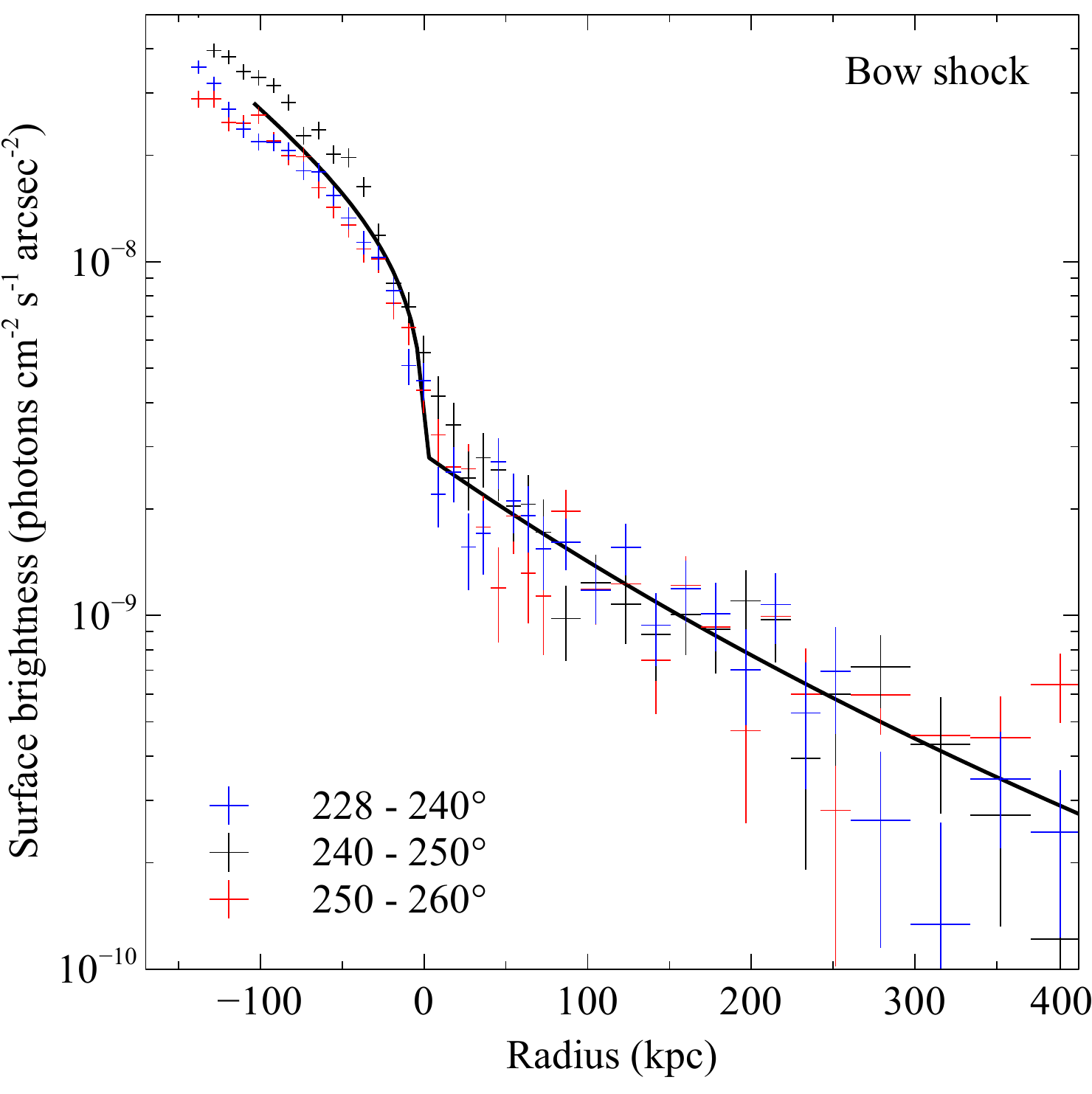}
\includegraphics[width=0.45\columnwidth]{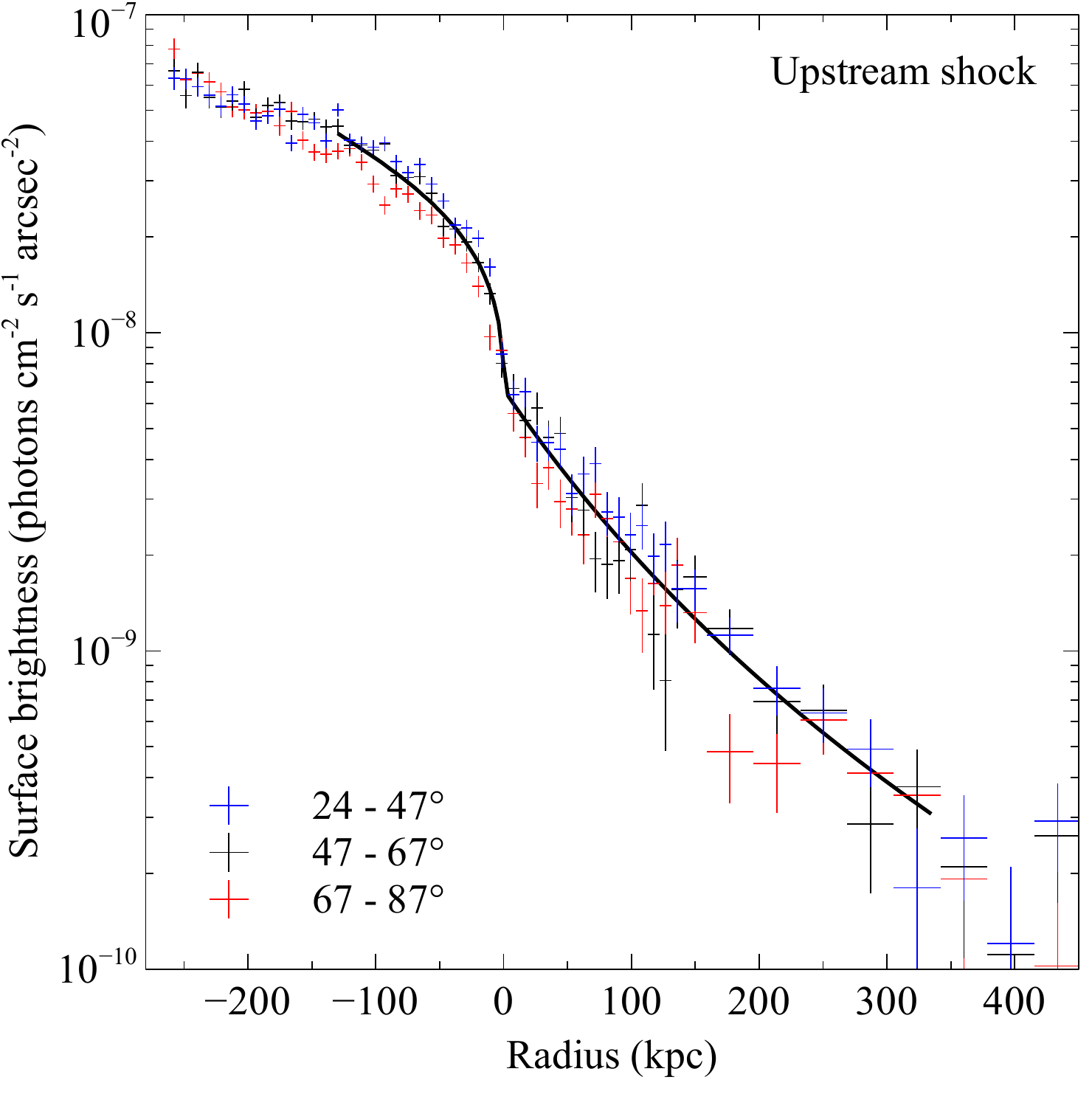}
\caption{Background-subtracted surface brightness profiles of the bow shock (left) and
  upstream shock (right) each divided into three sectors showing the
  variation in the shock width and strength with position angle.  The
  best-fit projected discontinuity models generated for the full width
  of the shock fronts in section \ref{sec:shocks} are shown overlaid.}
\label{fig:shockwidths}
\end{minipage}
\end{figure*}

Fig. \ref{fig:shockwidths} shows the variation in the surface
brightness profile of each shock front with position angle.  The bow
shock appears to be slightly stronger in the central sector but the
lower number of counts in this sector, particularly in the preshock
region, increases the error on the density jump value and this is
therefore not a significant result.  Both the bow and upstream shock
fronts appear to be consistently narrow across their respective
lengths of $\sim500\kpc$ and $\sim440\kpc$.

\subsection{Spectral analysis of the shock fronts}

\begin{figure*}
\begin{minipage}{\textwidth}
\centering
\includegraphics[width=0.45\columnwidth]{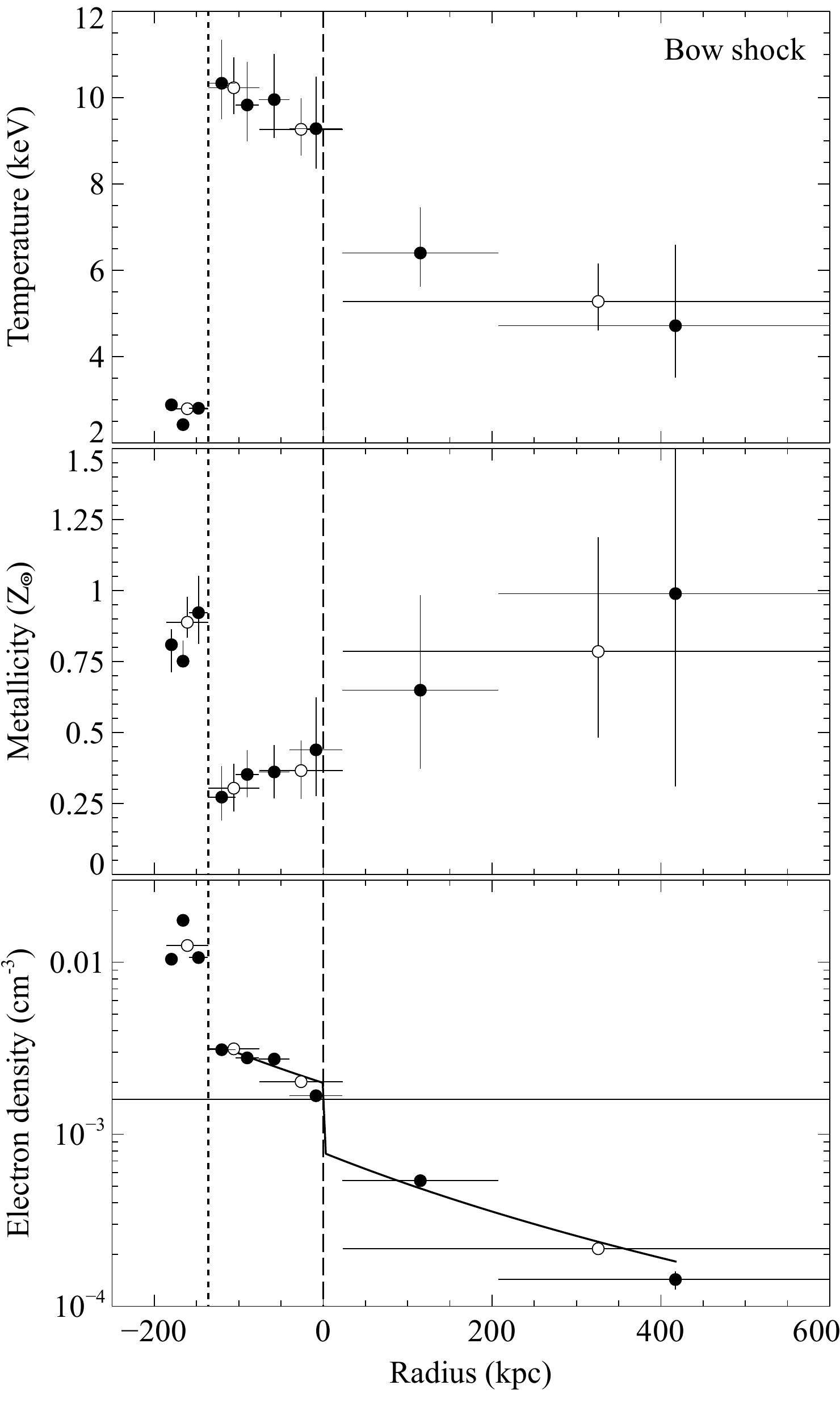}
\includegraphics[width=0.45\columnwidth]{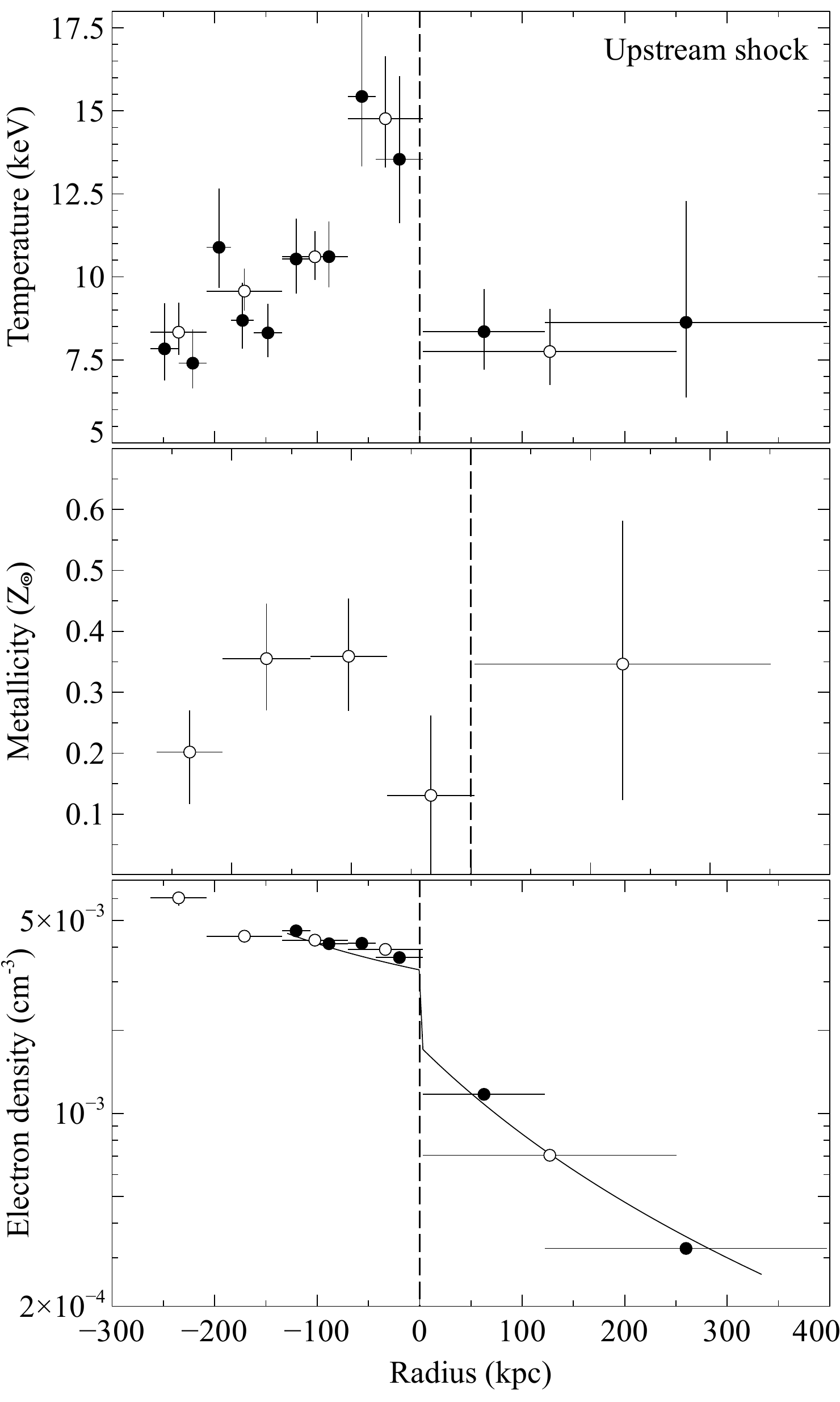}
\caption{Radial profiles through the bow (left) and upstream (right)
  shock sectors (shown in Fig. \ref{fig:SBsectors}) showing the
  projected electron temperature (upper), projected metallicity
  (centre) and deprojected electron density (lower).  There are two
  sets of points showing a finer (filled points) and a broader (open
  points) radial binning.  The electron density model (solid line)
  shown corresponds to the best-fit to the surface brightness profile
  across each shock (section \ref{sec:shocks}).  The vertical
  long-dashed line shows the location of each shock front and the
  vertical short-dashed line shows the position of the cold front on
  the leading edge of the subcluster core. }
\label{fig:paramprofiles}
\end{minipage}
\end{figure*}

The temperature jump across the shock front can be used to derive an
independent measure of the Mach number and to calculate the shock
velocity to determine the time since core passage of the subcluster.
Fig. \ref{fig:paramprofiles} shows the projected temperature,
projected metallicity and deprojected electron density profiles for
the bow shock and upstream shock sectors (see
Fig. \ref{fig:SBsectors}).  These radial profiles have the same radius
of curvature as each of the two shock fronts and therefore clearly
show the temperature and density decreases across the shock edges.  We
calculated the deprojected temperature profile across each shock front
using \textsc{projct} in \textsc{xspec}.  Deprojection routines, such
as \textsc{projct}, generally assume that the cluster is spherically
symmetric, which is a reasonable assumption for a relaxed cluster but
can be problematic for major mergers.  However, for the sectors across
the shock fronts, the cluster appears approximately circular on the
sky and the sharp increase in surface brightness across the edge
reduces the impact of projected emission.  The large size of the
radial bins used to measure the temperature values is likely to be a
more significant source of error in determining the temperature jump
across each shock.

We find that the deprojected temperature decreases by a factor of
$T_2/T_1=1.8\pm0.3$ at the bow shock and $T_2/T_1=2.1^{+0.4}_{-0.3}$
at the upstream shock.  The Rankine-Hugoniot shock jump conditions
directly relate the temperature jump to the density jump.  We can
therefore calculate the expected density jump and Mach number of each
shock (eq. \ref{eq:Mdensity}) from the observed temperature jump as an
independent verification (\citealt{LandauLifshitz59}).  We calculate a
Mach number $M=1.8^{+0.3}_{-0.2}$ for the bow shock and $M=2.0\pm0.3$
for the upstream shock.  The Mach number for both shock fronts is
therefore consistent within the errors for the calculations using both
the temperature and density jumps.  The temperature jump provides a
less accurate measure of the Mach number as the errors are greater
than for the density and larger radial bins must be used to measure
the temperature so we cannot resolve the jump across the shock accurately.


By using the bow shock speed and estimating the distance between the
subcluster core and the primary cluster centre, we calculate the time
since the subcluster passed through the primary cluster core.  The
shock speed, $v$, is calculated by multiplying the shock Mach number
and the sound speed in the preshock gas $c_s=(\gamma
k_BT_1/m_{\mathrm{H}}\mu)^{1/2}$, where $T_1$ is the preshock gas
temperature and $\mu=0.6$ is the mean molecular weight in the ICM.
For the bow shock in Abell 2146, we use the Mach number calculated
from the density jump $M=2.3\pm0.2$ and calculate the preshock sound
speed to be $c_s=1170^{+100}_{-70}\kmps$, giving a shock velocity
$v=2700^{+400}_{-300}\kmps$.  However, even in the new, deep
\textit{Chandra} observations, it is not possible to conclusively
identify an X-ray peak associated with the primary cluster core
(Fig. \ref{fig:maps}, upper left).  Simulations suggest that if the
primary cluster halo has a low concentration, the core will not
survive a major collision and the X-ray peak will be destroyed
(eg. \citealt{Mastropietro08}).  We have therefore assumed that the
collision between the two cluster cores occurred at a peak in the
pressure map between the subcluster tail end and the upstream shock
(Fig. \ref{fig:maps}, left).  The distance between the subcluster core
and the estimated location of the core collision site is then
$\sim350\kpc$ and the time since core passage is therefore
$\sim0.1-0.2\Gyr$.  This is only a rough estimate of the timescale as
the subcluster velocity is likely to be significantly lower than the
shock velocity (\citealt{Springel07}) and the location of the
collision site may be closer to the subcluster tail and the plume.
However, it is clear from the detection of the shock fronts and the SW
plume structure that this merger is observed soon after core
collision.

Fig. \ref{fig:paramprofiles} (left) also shows that there is a sharp
change in the gas properties across the cold front at the leading edge
of the subcluster cool core.  The temperature increases by a factor of
$3.7^{+0.3}_{-0.2}$ (discussed further in section
\ref{sec:conduction}) whilst the metallicity drops from
$0.89^{+0.09}_{-0.05}\Zsun$ inside the subcluster core to
$0.30^{+0.09}_{-0.08}\Zsun$ in the postshock gas.  The sharp
metallicity drop clearly demonstrates the different origins of the ICM
on either side of the contact discontinuity.  The high metallicity
core from the subcluster is travelling through the ICM in the
outskirts of the primary cluster, which has a much lower metallicity.
In comparison, the upstream shock sector (Fig. \ref{fig:paramprofiles}
right) has an approximately constant metallicity of $0.3-0.4\Zsun$, as
it consists predominantly of primary cluster gas.  There may be more
metal-enriched ram pressure stripped material from the subcluster in
this sector, however the material that was stripped from the core
earlier is likely to have a lower metallicity than the current peak
and the subcluster tail shows no evidence of a strong metallicity
gradient (Fig. \ref{fig:maps}).

\subsection{The establishment of electron-ion equilibrium}
\label{sec:electronion}

\begin{figure}
\centering
\includegraphics[width=0.95\columnwidth]{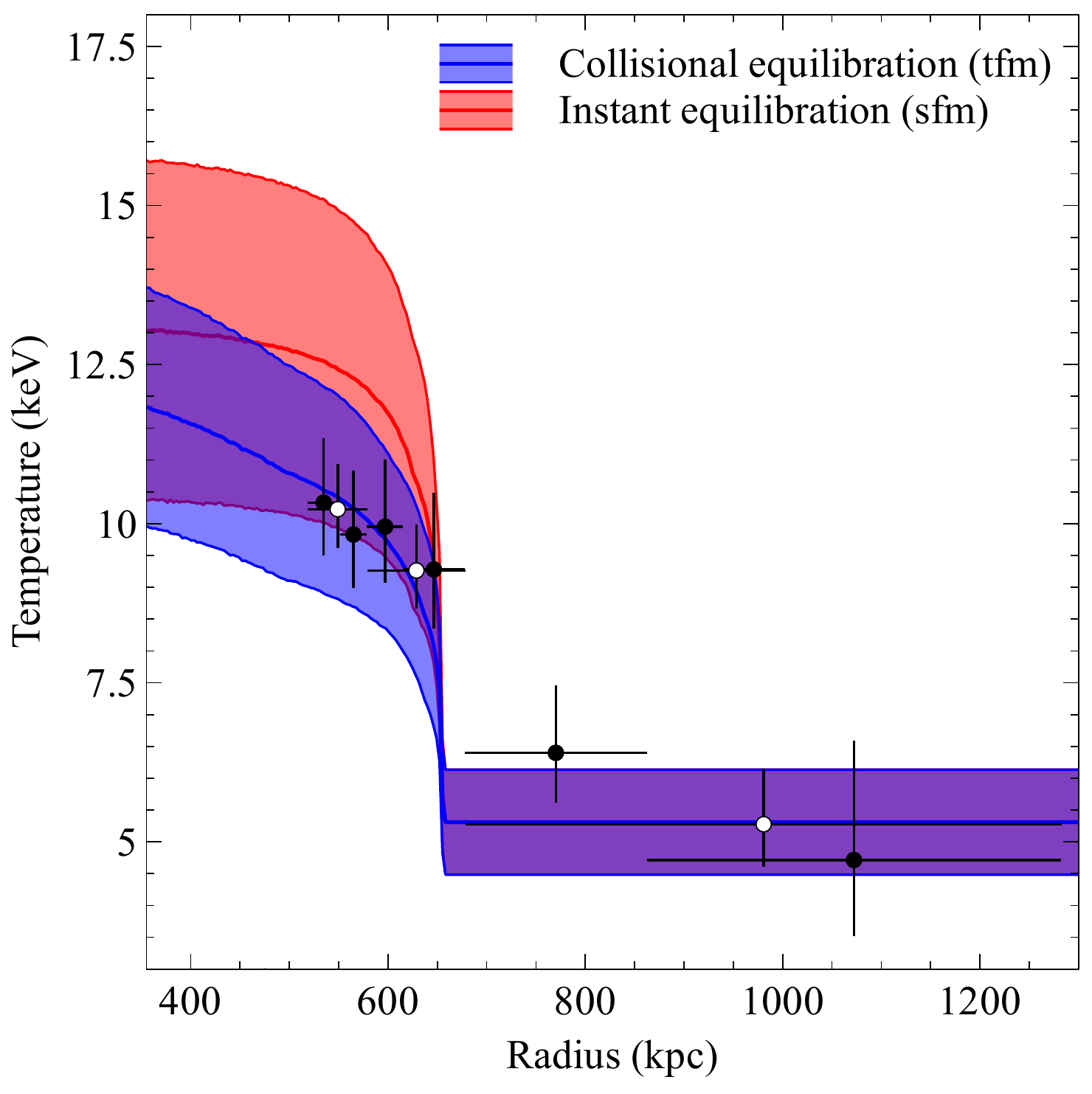}
\caption{The re-establishment of electron-ion equilibrium behind the
  bow shock.  The projected electron temperature profile is overlaid
  with model predictions (with $1\sigma$ error bands) for instant
  equilibration (red) and adiabatic compression followed by
  collisional equilibration (blue).  The open and filled data points
  show narrower and broader binning of the temperature profile.}
\label{fig:bowelectronion}
\end{figure}

\begin{figure}
\centering
\includegraphics[width=0.95\columnwidth]{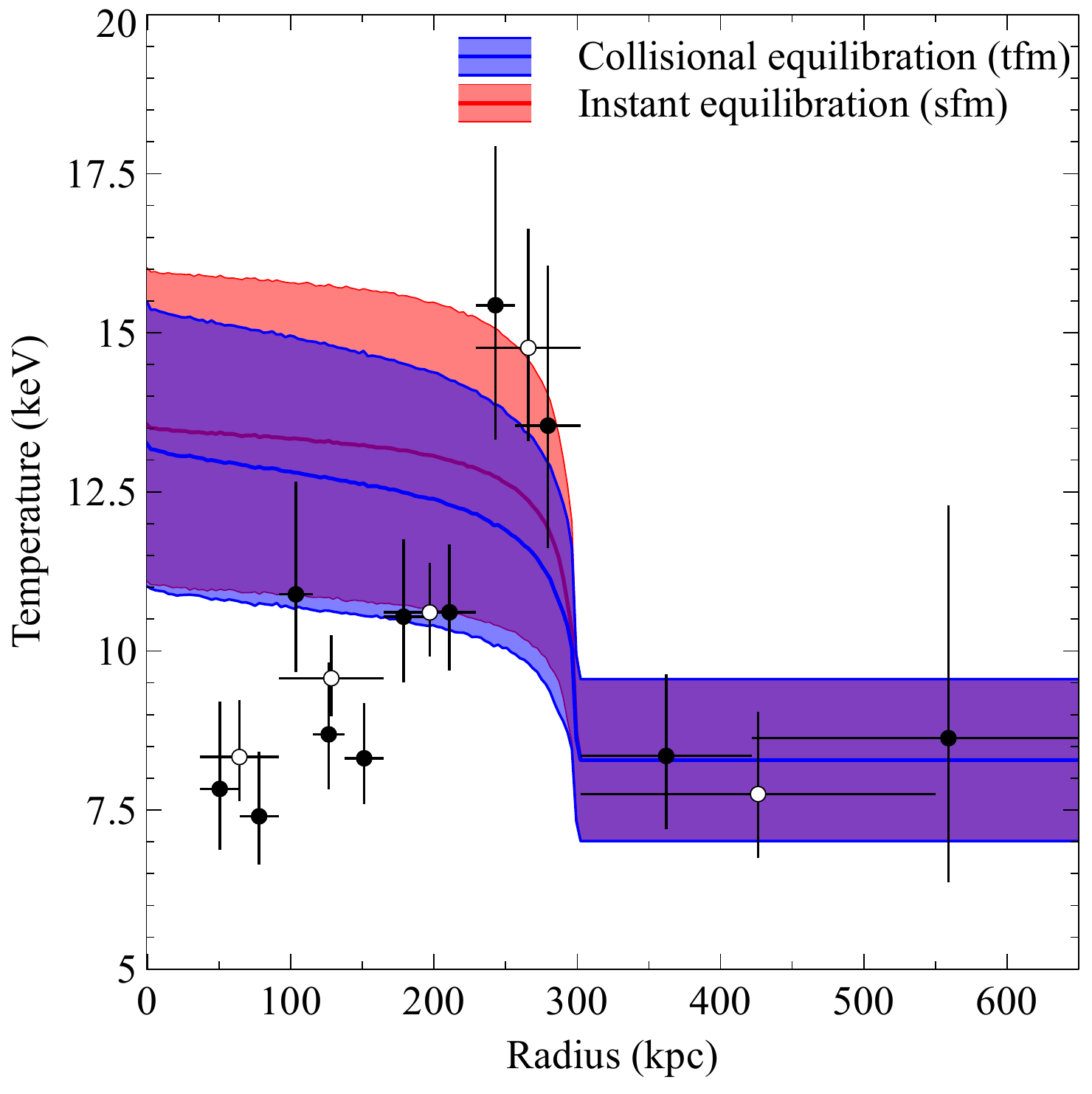}
\caption{The re-establishment of electron-ion equilibrium behind the
  upstream shock.  The projected electron temperature profile is
  overlaid with model predictions (with $1\sigma$ error bands) for
  instant equilibration (red) and adiabatic compression followed by
  collisional equilibration (blue). The open and filled data points
  show narrower and broader binning of the temperature profile.}
\label{fig:upstreamelectronion}
\end{figure}

Cluster merger shock fronts present a unique opportunity to
investigate the electron-ion equilibration time in the magnetized
ICM by mapping the postshock electron temperature.  The
Rankine-Hugoniot shock jump conditions can be used to calculate the
postshock temperature for the ICM electrons and ions once they reach
equilibrium after the passage of the shock
(eg. \citealt{LandauLifshitz59}).  However, the fraction of a shock's
kinetic energy that is initially transferred to the thermal and cosmic-ray
populations of the electrons and ions remains an open question.  

A shock front propagating through a collisional plasma heats the ions
dissipatively in a shock layer that has a width of order the
collisional mean free path.  The electrons have a much higher thermal
velocity and are not strongly heated by merger shocks.  They are
compressed adiabatically and subsequently equilibrate with the ions
according to the Coulomb collisional timescale (\citealt{Spitzer62})
given by,

\begin{equation}
t_{eq}\mathrm{(e,p)}\approx 6.2\times10^{8}\yr
\left(\frac{T_e}{10^8\K}\right)^{3/2}\left(\frac{n_e}{10^{-3}\pcmcu}\right)^{-1}
\label{eq:equilib}
\end{equation}

\noindent where $T_e$ is the electron temperature and $n_e$ is the electron
density (see eg. \citealt{Sarazin88}).  

However, shocks in a magnetised plasma, such as the ICM, are likely to
be collisionless.  Observations of the solar wind shocks found that
the electron and ion temperature jump occurs in a shock layer several
orders of magnitude thinner than the mean free path
(eg. \citealt{Ness64}; \citealt{Montgomery70}; \citealt{Hull01}).  The
coupling of particles to electric and magnetic fields produces
interactions which have dissipation scale lengths much shorter than
the ordinary collision mean free path.  The plasma waves producing
these interactions and instabilities affect ions and electrons
differently due to the large difference in mass (for a review see
eg. \citealt{Tidman71}; \citealt{Friedman71}).  We might therefore
expect to find an electron heating rate shorter than the Coulomb
collisional timescale behind a cluster merger shock.

Measurements of the postshock temperatures in supernova remnants
(eg. \citealt{Rakowski05}; \citealt{Raymond05}; \citealt{Ghavamian07})
and heliospheric shocks (eg. \citealt{Schwartz88};
\citealt{Russell05}) show that in most cases the electrons are heated
less than the protons, $T_{e}/T_{i}<1$, in regions close to the shock
front.  However, these measurements cannot determine the timescale of
subsequent equilibration, $t_{eq}\mathrm{(e,p)}$, as this corresponds
to a linear scale of several A.U. for planetary bow shocks and is
comparable to the age of the remnant for supernova shocks.  For cluster
merger shock fronts the equilibration timescale corresponds to a
distance of hundreds of kpc, so while clusters are too distant for us
to resolve the shock layer, we can study the subsequent equilibration
of the ICM constituents.

The bow shock in the Bullet cluster provided the first measurement of
the electron-ion equilibration timescale in the ICM.
\citet{Markevitch06} compared the observed electron temperature
profile across the shock front with two models for equilibration.  The
instant equilibration model predicts that the electrons are strongly
heated at the shock front and the electron temperature rapidly
increases to the postshock temperature, similar to the ion
temperature.  The collisional model predicts an adiabatic compression
of the electron population at the shock and a subsequent slower
equilibration with the ions on a timescale determined by Coulomb
collisions.  The observed temperature profile for the Bullet cluster
supported instant equilibration, indicating that electrons were
rapidly heated at the shock front on a timescale faster than Coulomb
collisions.  However, the postshock temperature in the Bullet cluster
is very high ($\sim20-40\keV$) compared to the \textit{Chandra} energy
band and therefore difficult to constrain.


Although the postshock temperatures in Abell 2146 are lower than the
Bullet cluster, the shock Mach numbers are also lower, which reduces
the separation between the collisional and instant equilibration
models.  Figs. \ref{fig:bowelectronion} and
\ref{fig:upstreamelectronion} show the observed and model projected
electron temperature profiles behind the bow shock and upstream shock,
respectively.  The observed profiles were generated using single
temperature model fits to spectra extracted from two different sets of
radial bins.  The narrower radial bins were selected to contain a
minimum of 2000 counts and the broader bins have a minimum of 4000
counts.  The postshock radial bin closest to the shock boundary was
positioned to overlap the fainter preshock gas and ensure minimal
contamination of the neighbouring preshock gas radial bin.  

The models were generated by using the observed preshock electron
temperature and the density jump in the Rankine-Hugoniot shock jump
conditions to predict the postshock gas temperature.  For the instant
equilibration model (a single fluid model), the electron temperature
and the ion temperature jump at the shock front to the postshock gas
temperature predicted by the shock jump conditions.  This temperature
model was then projected along the line of sight, using the best-fit
surface brightness model determined for each shock front in section
\ref{sec:shocks}, to produce the final projected temperature model.
The projection smoothes the sharp jump in temperature at the shock
front.  Although the projection assumes a constant preshock
  temperature, the steep decline in the surface brightness across and
  ahead of each shock front (Fig. \ref{fig:SBprofiles}) meant the
  effect of any cooler gas beyond the observed temperature points is
  negligable.

For the collisional equilibration model (a two fluid model), the
electron temperature increases at the shock front according to the
adiabatic compression of the particles,

\begin{equation}
T_{e,2} = T_{e,1}\left(\frac{\rho_{2}}{\rho_{1}}\right)^{\gamma-1}.
\end{equation}

\noindent Coulomb collisions then subsequently equilibrate the
electron and ion temperatures at a rate

\begin{equation}
\frac{\mathrm{d}T_e}{\mathrm{d}t} = \frac{T_i - T_e}{t_{eq}}.
\label{eq:equilibrate}
\end{equation}

\noindent We require that the total kinetic energy density is
conserved so that the local mean gas temperature is given by

\begin{equation}
T_{\mathrm{gas}}=\frac{n_iT_i + n_eT_e}{n_{\mathrm{gas}}}=\frac{T_i+1.1T_e}{2.1}.
\label{eq:conserv}
\end{equation}

\noindent The mean gas temperature is constant and can be calculated
from the Rankine-Hugoniot shock jump conditions.  Therefore, although
X-ray observations do not currently have the spectral resolution
required to measure the ion temperature directly, we can use this
requirement to determine the immediate postshock ion temperature.  The
postshock ion temperature is significantly higher than the
adiabatically-compressed electron temperature and for the collisional
model we assume that the electrons equilibrate with the hotter ions
according to the Coulomb collisional timescale, $t_{eq}$.  Using
eq. \ref{eq:equilib} we calculate this timescale for the postshock
electrons.  By multiplying $t_{eq}$ by the shock velocity in the
postshock gas (bow shock $v_{ps} = 1100\pm100\kmps$, upstream shock
$v_{ps} = 1300\pm200\kmps$), we determine the distance behind the
shock where equilibration is reached.  Electrons at this position were
heated by the shock $t_{eq}$ years ago.  By integrating
eq. \ref{eq:equilibrate}, and using eqs. \ref{eq:equilib} and
\ref{eq:conserv}, the electron temperature as a function of distance
behind the shock was determined analytically (see
eg. \citealt{Fox97}; \citealt{Ettori98}).  This collisional model was then also
projected along the line of sight to produce the final model for
comparison with the observed projected electron temperature.

The instant and collisional models were projected by determining the
model electron temperature in small volumes, $\mathrm{d}V$, along the
line of sight for a particular annulus and calculating the
corresponding emission measure using the best-fit density
discontinuity model (section \ref{sec:shocks}).  For each annulus, the
emission measures for $\mathrm{d}V$ with similar temperatures were
summed together using a set temperature binning with fine resolution
($\sim0.1\keV$).  To determine the projected temperature in this
annulus, we produced fake spectra in \textsc{xspec} using
multi-component \textsc{mekal} models with the temperature of each
component set to the midpoint of each temperature bin and the
normalization set to the corresponding summed emission measure.  The
metallicity was fixed to $0.4\Zsun$ and responses for the appropriate
detector region were used.  The \textsc{mekal} model components were
also combined with a \textsc{phabs} absorption component set to the
Galactic column density.  These fake spectra were then fitted with
single absorbed \textsc{mekal} models, with fixed metallicity and
column density, to determine the final model projected temperature.


The main sources of error for these models were the measurement of the
preshock temperature and, to a lesser extent, the density jump and
preshock electron density.  We used a Monte Carlo technique to
determine the uncertainties in the model projected temperature profiles.  We
repeated the model calculation and projection 1000 times, each time
using new values of the preshock temperature, density jump and
preshock density based on Gaussian distributions.  The output models
are the median of this process and the $1\sigma$ errors are calculated
from the 15.85 and 84.15 percentile spectra.

Fig. \ref{fig:bowelectronion} shows the observed and model projected
electron temperature profiles across the bow shock.  The observed
temperature profile is cut off $\sim150\kpc$ behind the bow shock
front where the temperature drops in the subcluster core.
Unfortunately, owing to the large uncertainty on the preshock count
rate in the earlier $45\ks$ observation, there were fewer preshock
counts in the new $400\ks$ observation than anticipated.  The lower
number of counts produced a greater than expected error on the crucial
preshock temperature, which is the main source of uncertainty for the
models, and therefore it is not possible to conclusively exclude
either model.  However, the observed postshock electron temperatures
for the bow shock appear lower than predicted by the instant
equilibration model and favour the collisional equilibration model.
We therefore conclude that collisional equilibration cannot be ruled
out for cluster merger shocks.

The preshock temperature could potentially be higher than the
conservative estimate of $5.3^{+0.9}_{-0.7}\keV$ used in this
analysis.  The temperature bin closest to the bow shock front has a
slightly higher temperature of $6.4^{+1.1}_{-0.8}\keV$.  This increase
in temperature could be due to contamination from the higher
temperature and density postshock region.  However, the neighbouring
postshock temperature bin was positioned to significantly overlap the
preshock region by $20\kpc$ ($6\asec$) and minimise this effect.  It is
therefore more likely that there is a gradient in the preshock gas temperature
which would increase the postshock temperature of both models and
strengthen our conclusion that collisional equilibration is possible
for cluster merger shocks.

Measurement of the equilibration timescale was more difficult for the
upstream shock as the Mach number was lower, reducing the separation
between the instant and collisional models.  In addition, the
postshock electron temperatures are $\sim5\keV$ higher compared to the
bow shock, which increased the uncertainty on the values.
Fig. \ref{fig:upstreamelectronion} shows that the electron temperature
increases rapidly in the postshock region to $15^{+2}_{-1}\keV$ and
then drops to $\sim10\keV$ approximately $100\kpc$ behind the shock
front.  Although the errors are larger, the electrons behind the
upstream shock appear to equilibrate faster than those at the bow
shock.  However, the upstream shock is also propagating through the
primary cluster outskirts, which have been disturbed by the
subcluster's passage.  There is likely to be significant substructure
and additional shock heating in the region of the upstream shock
front.  Ram pressure stripped material from the subcluster tail is
likely to be the cause of the rapid decline in temperature
$\sim70\kpc$ behind the shock.  We therefore place less weight on the
conclusions on equilibration behind the upstream shock as the
situation here is more complex and the errors on the electron
temperature are greater.



\subsection{Non-equilibrium ionization}
\label{sec:nei}

In the previous section, we have shown that the postshock ICM
significantly deviates from thermal equipartition between electrons
and ions but our analysis has so far assumed that the ICM is in
ionization equilibrium.  However, while the temperature of the ICM
increases rapidly in the shock layer, the ionization state of the ions
still reflects the preshock temperature and the ICM will be
underionized compared to the equilibrium case.  The ionization balance
will be recovered by collisions on a timescale of $\sim10^{7}\yr$ for
an electron density $\sim10^{-3}\pcmcu$ but until this is achieved
there will be more ionizations than recombinations.  Simulations of
the outer regions of clusters and of merging systems have considered
the effects of non-equilibrium ionization, in particular the effect on
the intensity ratios of the Fe\,K\,$\alpha$ lines
(eg. \citealt{Yoshikawa06}; \citealt{Akahori10,Akahori11};
\citealt{Wong11}).  We have therefore considered whether
non-equilibrium ionization will produce an observable effect in the
postshock ICM of Abell 2146.

For a temperature of $\sim9\keV$ ($10^{8}\K$) behind the bow shock,
the typical density-weighted timescale for the ions to approach
ionization equilibrium is $\tau\sim4\times10^{12}\pcmcus$ for Fe and
significantly shorter for other important ions in the ICM
(eg. \citealt{Smith10}).  Using the postshock electron density,
$1.99\pm0.09\times10^{-3}\pcmcu$, and the velocity of the shock in the
postshock gas (section \ref{sec:electronion}), we estimate that the Fe ions
reach equilibrium ionization $\sim50\kpc$ behind the bow shock.
Therefore, only the first narrow radial bin behind the bow shock (and
also for the upstream shock) could be affected by non-equilibrium
effects.

We fit each of these radial bins, immediately behind each shock front,
with an absorbed non-equilibrium ionization \textsc{nei} spectral
model in \textsc{xspec} (eg. \citealt{Hamilton83};
\citealt{Borkowski01}).  The \textsc{nei} model indicated a
density-weighted ionization timescale of $\tau>4\times10^{13}\pcmcus$
behind each shock front suggesting that there was no measurable
signature of non-equilibrium ionization in Abell 2146.  This is a
difficult measurement given the low spectral resolution at high
energies, covering the Fe~\textsc{xxv}\,K\,$\alpha$ and
Fe~\textsc{xxvi}\,K\,$\alpha$ lines (\citealt{Akahori10,Akahori11}),
and the low photon count rates at the shock front, necessitating large
radial bins relative to the ionization equilibration timescale.



\begin{figure*}
\begin{minipage}{\textwidth}
\centering
\includegraphics[width=0.45\columnwidth]{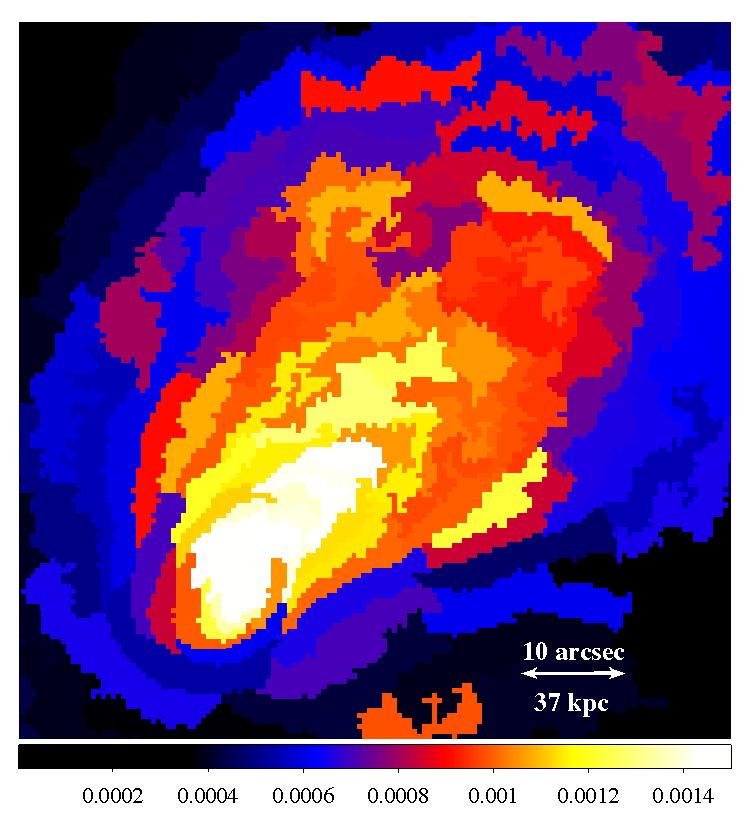}
\includegraphics[width=0.45\columnwidth]{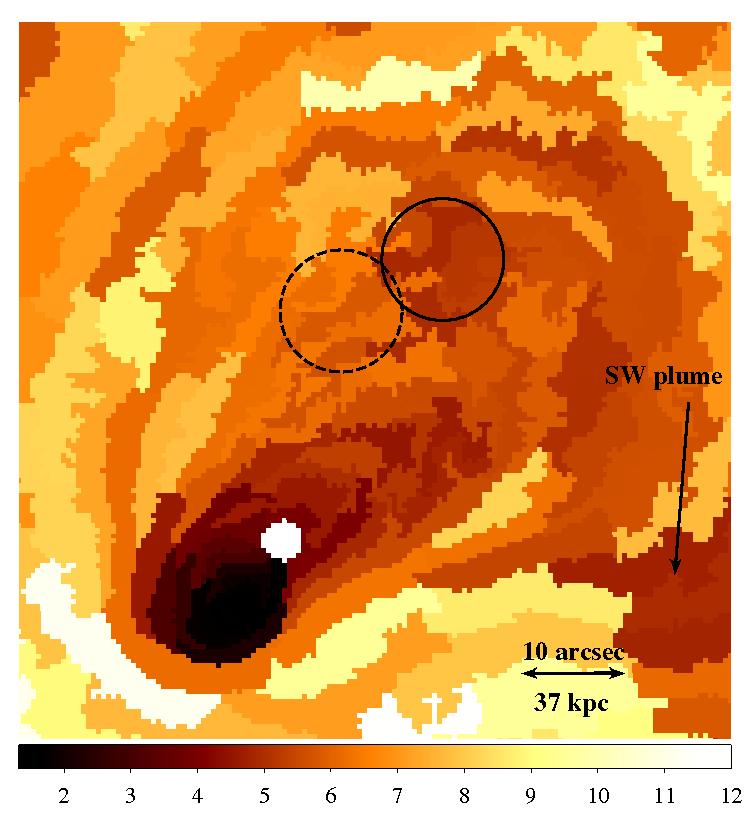}
\caption{Left: projected pseudo-pressure map of the subcluster tail
  with $S/N\geq32$ in each bin (units $\pseudoP$).  Right: projected
  temperature map (units keV) of the same region with $S/N\geq32$.
  The excluded central AGN is visible as a small white circle.  The
  circles mark a region where the temperature gradient through the
  subcluster tail reverses from $6-6.5\keV$ (dashed) to $5\keV$ gas
  (solid).}
\label{fig:tailmaps}
\end{minipage}
\end{figure*}
\begin{figure*}
\begin{minipage}{\textwidth}
\centering
\includegraphics[width=0.45\columnwidth]{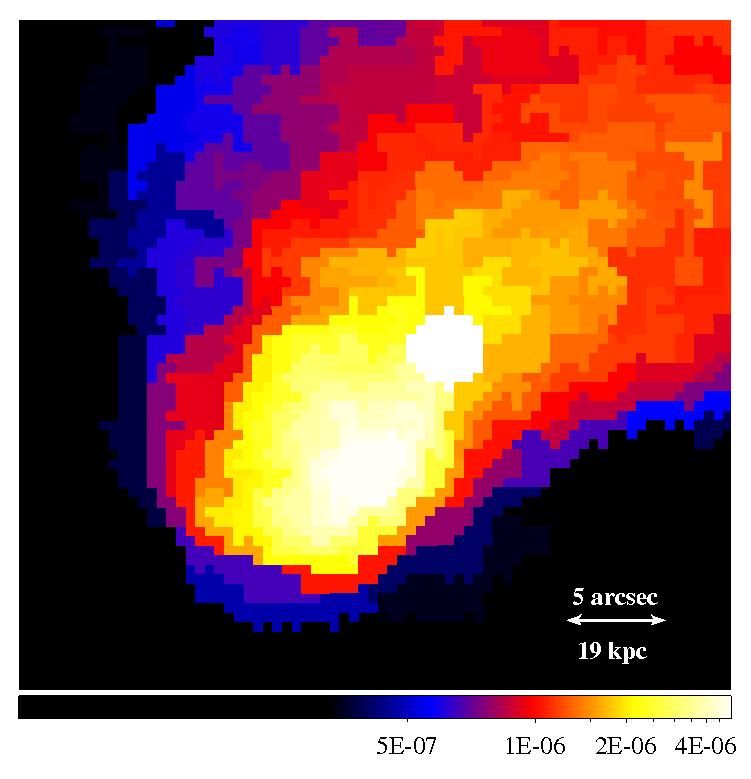}
\includegraphics[width=0.45\columnwidth]{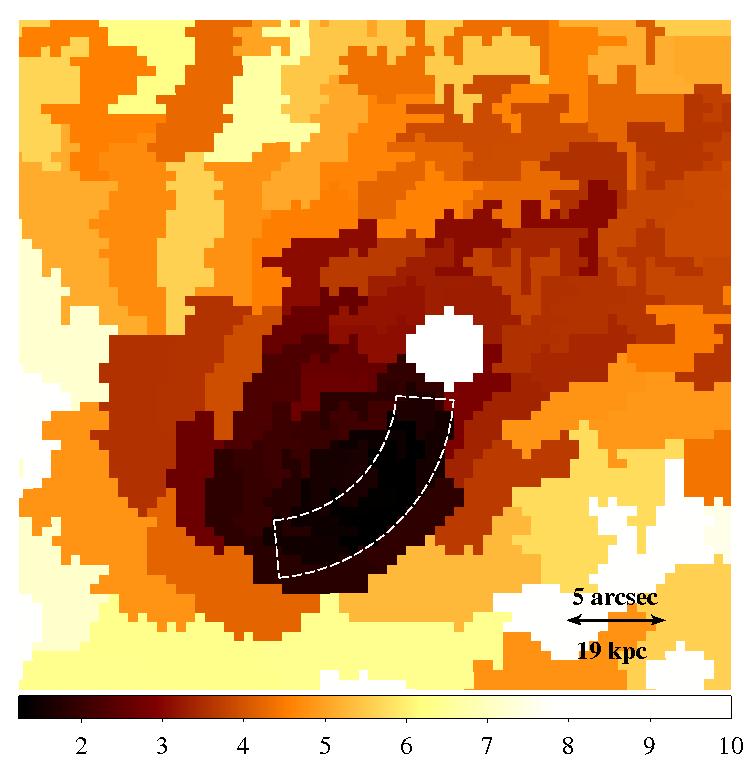}
\caption{Left: projected emission measure per unit area map of the
  subcluster core with $S/N\geq15$ in each bin (units $\empasecsq$,
  see Fig. \ref{fig:maps}).  Right: projected temperature map (units
  keV) of the same region with $S/N\geq22$.  The white dashed region
  marks the cold gas filament.  The excluded central AGN is visible as
  a small white filled circle in both images. }
\label{fig:coremaps}
\end{minipage}
\end{figure*}

\section{Disruption of the subcluster core}
\label{sec:core}

Fig. \ref{fig:tailmaps} shows the detailed structure in the subcluster
core and ram pressure stripped tail.  The front edge of the subcluster
core appears roughly spherical but the sides are strongly sheared by
the surrounding shocked gas.  The southern edge of the subcluster core
appears to be narrow with no sign of the disruption and stripping
which affects the eastern and northern edges
(eg. Fig. \ref{fig:mainimages}).  The temperature gradients on the SW
and NE sides of the subcluster core are very different.  There is
clearly a more graduated increase in temperature along the NE edge
where the core is breaking up.

Fig. \ref{fig:coremaps} shows that the filament of gas running along
the southern edge of the core, and apparently ending at the AGN, contains the
coldest gas in the cluster.  Using a single temperature spectral model
(section \ref{sec:spectro}), we determined that the X-ray gas
temperature drops to $1.48^{+0.09}_{-0.08}\keV$ in this filament.
This cool X-ray gas filament is also detected as a narrow, coherent
filament in H$\alpha$ and [N~\textsc{ii}] observations
(\citealt{Canning11}).  We fitted the spectrum from this region
(shown in Fig. \ref{fig:coremaps}) with a multi-temperature model to
determine if there is a significant amount of multi-phase gas
detectable in the X-ray.  Following \citet{Sanders04}, we used
multiple absorbed \textsc{mekal} components with temperatures fixed at
0.5, 1, 2 and $4\keV$, common metallicities and free normalizations.
We tested an additional $8\keV$ component but found that the best-fit
normalization was consistent with zero.  The majority of the gas in
the filament is at around $2\keV$ but we detect significant amounts of
cooler gas.  The fraction of the emission measure of gas at $1\keV$,
with respect to the total emission measure, is $15-20$\% and at
$0.5\keV$ is $5-10$\%.  Both of these lower temperature components are
detected at above $3\sigma$.  The best-fit metallicity for this
multi-temperature model is $0.8\pm0.1\Zsun$.  It is therefore likely
that the low metallicity observed in the subcluster core in
Figs. \ref{fig:maps} and \ref{fig:sectorprofiles} is a bias caused by
the use of only a single temperature model (\citealt{Buote94};
\citealt{Buote00}).

Using an absorbed \textsc{mekal + mkcflow} model, with the lower
temperature of the \textsc{mkcflow} component (\citealt{Mushotzky88})
fixed to $0.1\keV$, the upper temperature component tied to the
temperature of the \textsc{mekal} model and the metallicities tied
together, we determine an upper limit on the mass deposition rate for
the filament region of $40\Msunpyr$, which is cooling out of the X-ray
band (see eg. \citealt{McNamara06}; \citealt{Rafferty06}).  The upper
limit on the mass deposition rate for the whole of the subcluster core
is $50\Msunpyr$, therefore the vast majority of the cooling in the
subcluster core is likely occurring in this filament.  This is
significantly less than the star formation rate (SFR) of $192\Msunpyr$
determined from \textit{Spitzer} observations of the BCG by
\citet{ODea08}.  However the IR emission was not corrected for an AGN
contribution, which is likely to dominate (\citealt{Canning11}), and
therefore this SFR is an upper limit and the true rate is likely to be
much lower.


\subsection{Stripping of the NE edge}

Fig. \ref{fig:coremaps} shows the stripping of cool material from the
eastern edge of the subcluster cool core.  Cool gas is being pulled off the
eastern edge of the core into a $12\arcsec$ ($45\kpc$) filament which has a
steadily increasing temperature from $1.9\pm0.1\keV$ to
$3.1^{+0.5}_{-0.4}\keV$ along its length, although this could be
affected by hotter material seen in projection.  There are several
cooler blobs of gas to the North and NW of the AGN suggesting this
material then breaks off and may be directed West into the subcluster's
wake.  Warmer $3-4\keV$ gas appears to be filling underneath the cool,
stripped filament.  The coolest gas in the subcluster core is towards
the leading southern edge of the core.

Simulations of ram pressure stripping of
a cool core during a cluster merger show strong shearing of the
contact discontinuity and the formation of a strong vortex inside the
core, which transports material inside the core to the leading edge
and then back along the surface travelling with the surrounding flow
(eg. \citealt{Murray93}; \citealt{Balsara94}; \citealt{Heinz03}).  The
cooler blobs of gas stripped off the eastern edge trace the flow of material
around the subcluster core and show the flow converges behind the
core, at the approximate position of the AGN (see section
\ref{sec:specradial}).

In section \ref{sec:specradial} we suggest the subcluster is likely to
have passed to the North of the primary cluster core, colliding with a
small but non-zero impact parameter.  This trajectory will produce a
more curved orbit for the subcluster core, trending from a SE
direction to southern, compared to the case of the head on collision.  The southern
edge is now the leading edge of the subcluster core and there is
likely to be a greater shearing effect on the eastern edge due to the curved
trajectory.

If ram pressure stripping is the dominant mechanism disrupting the eastern
edge, we expect most of the cool core gas will be removed at a radius which
satisfies the condition

\begin{equation}
\rho_{\mathrm{s}}v_{\mathrm{rel}}^{2}\apprge P_{\mathrm{c}}(r),
\end{equation}

\noindent where $\rho_{\mathrm{s}}$ is the density of the surrounding
ambient ICM, $v_{\mathrm{rel}}$ is the relative velocity of the
ambient ICM and the subcluster cool core, and $P_{\mathrm{c}}$ is the
pressure profile of the subcluster core (eg. \citealt{Fabian91};
\citealt{Gomez02}).  The non-zero impact parameter of the merger makes
it difficult to determine the velocity of the ambient ICM around the
eastern edge of the subcluster.  We therefore estimate the maximum
velocity to be the postshock gas velocity of $1100\pm100\kmps$.  For
$n_e=0.0064\pm0.0002\pcmcu$ and
$P_{\mathrm{c}}=0.074\pm0.006\keVpcmcu$, we find that the condition
for ram pressure stripping is only approximately met along the eastern
edge of the core although the pressure does not steeply increase here
as observed inside the southern edge (Fig. \ref{fig:corecomp}, lower
right).  Therefore it's likely that the removal of material along this
edge is also facilitated by developing Kelvin-Helmholtz instabilities
(\citealt{Nulsen82}; \citealt{Inogamov99}).  However, without a more
effective estimate of the velocity of the ambient ICM around the
subcluster core it is difficult to calculate the timescale of the
developing instabilities and determine whether the subcluster core
will be subsequently destroyed (eg. \citealt{VikhlininBfield01};
\citealt{Heinz03}).

\subsection{Width of the subcluster southern edge}
\label{sec:conduction}

\begin{figure*}
\begin{minipage}{\textwidth}
\centering
\raisebox{1cm}{\includegraphics[width=0.4\columnwidth]{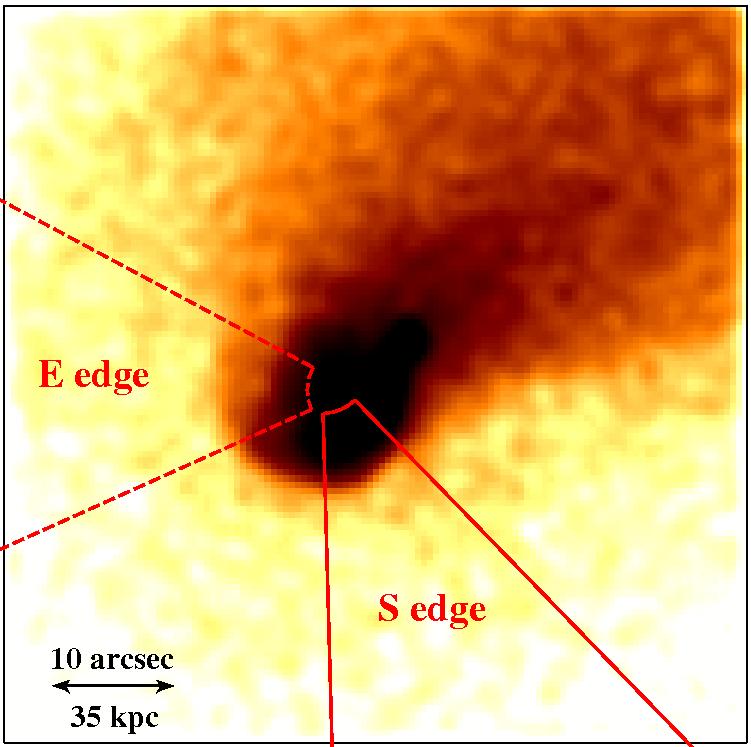}}
\includegraphics[width=0.45\columnwidth]{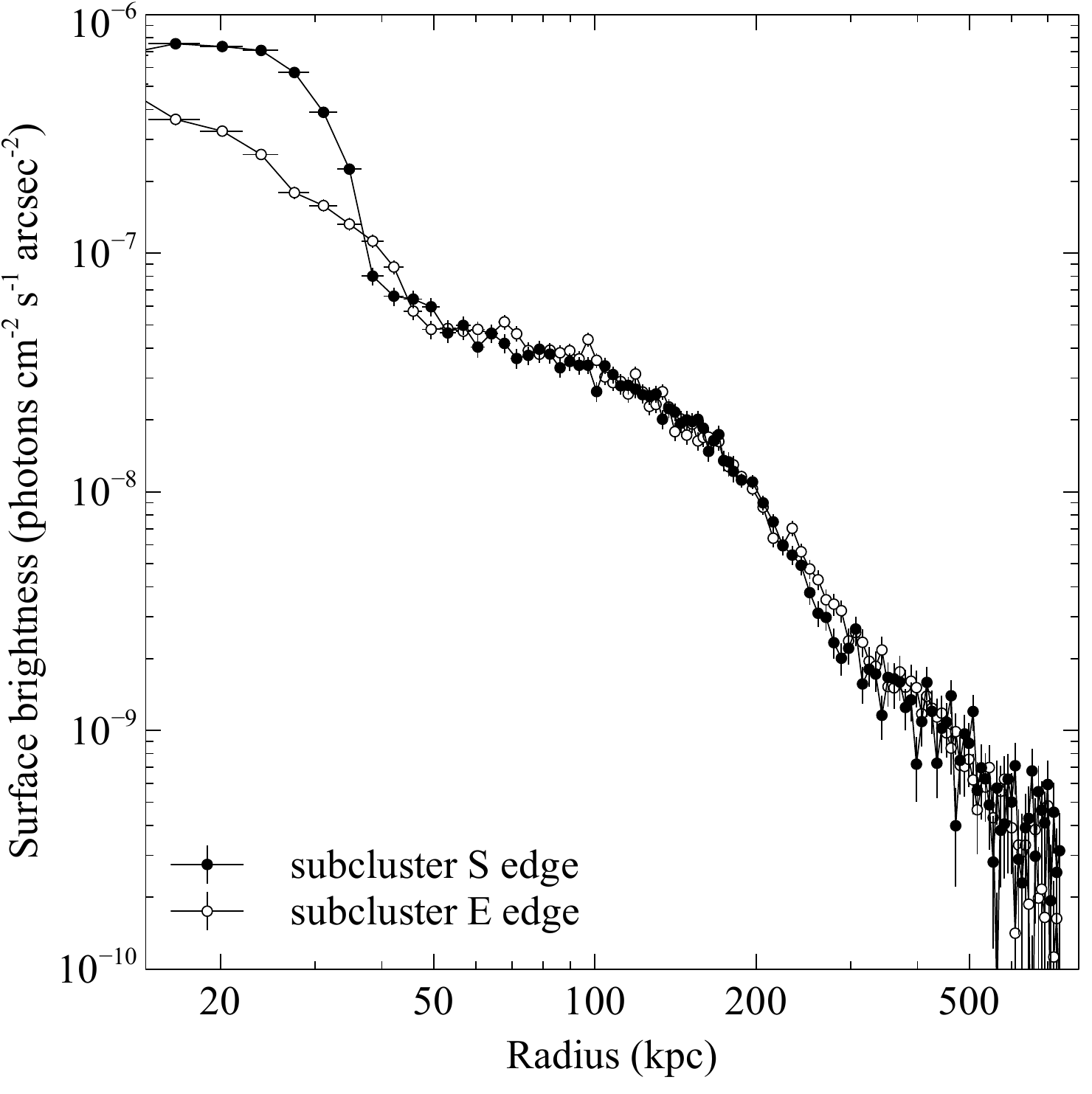}
\includegraphics[width=0.45\columnwidth]{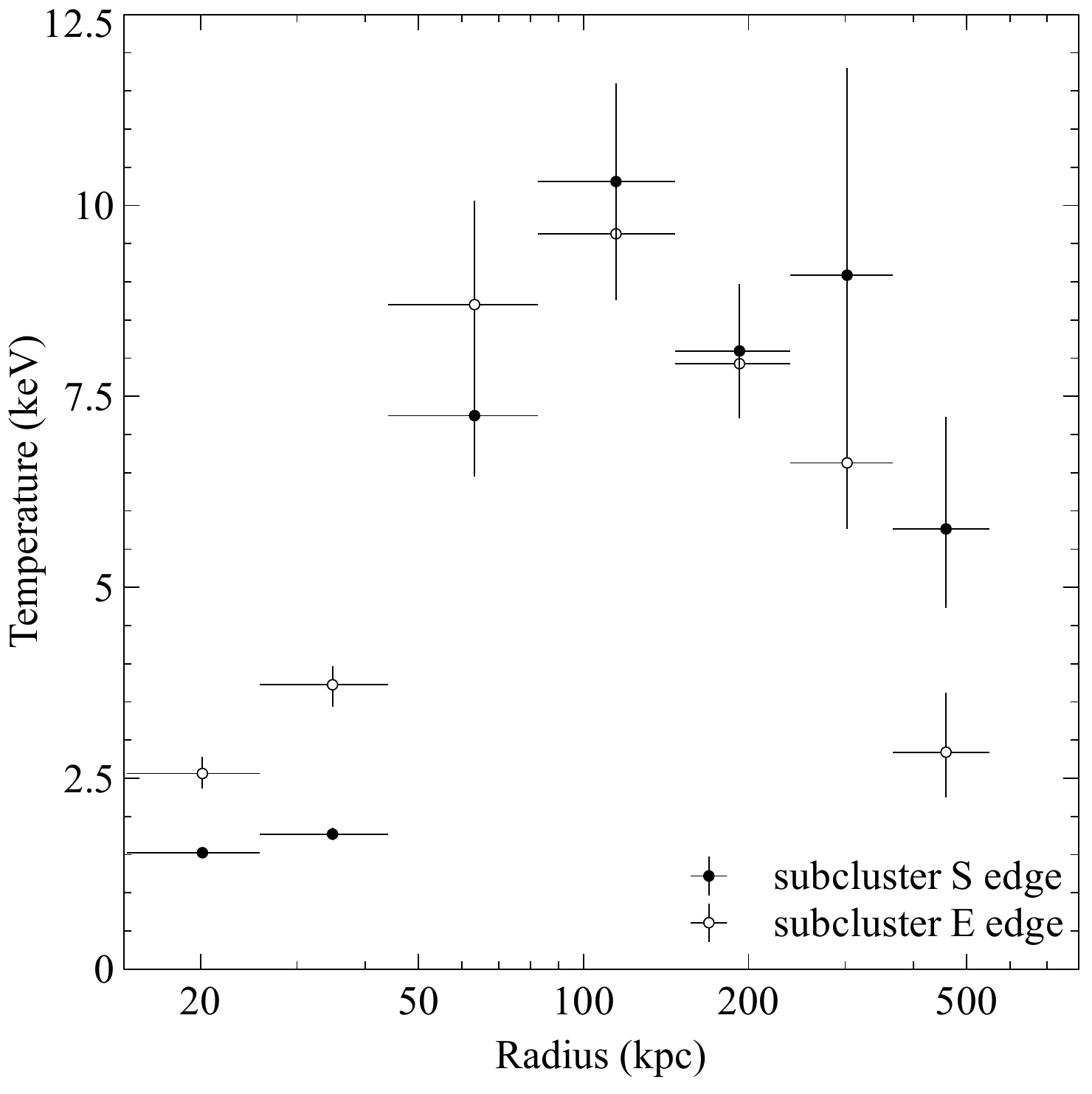}
\includegraphics[width=0.45\columnwidth]{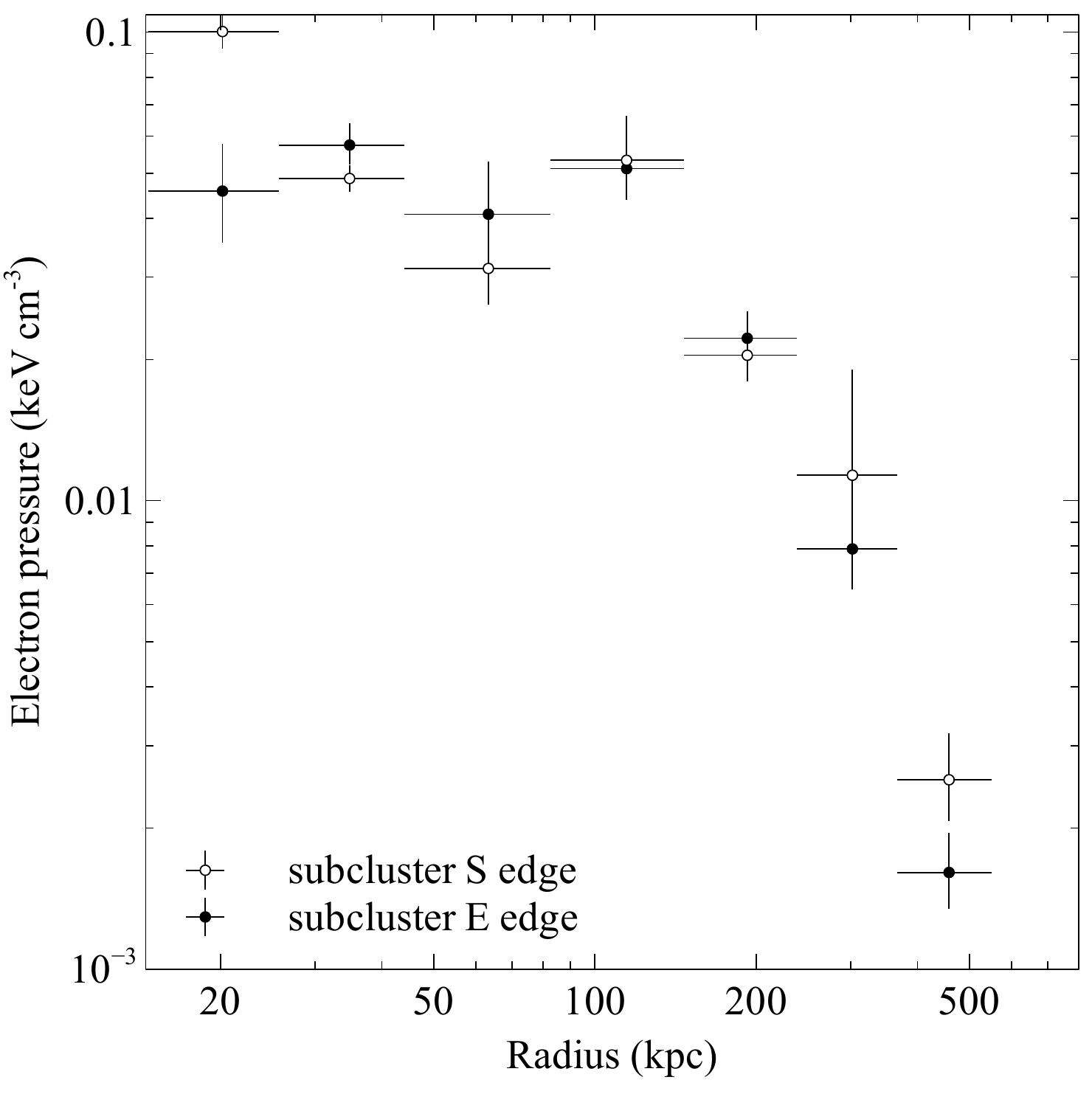}
\caption{Surface brightness (upper right), projected temperature
  (lower left) and electron pressure profiles (lower right) for two
  sectors across the southern (filled circles, $270-315^{\circ}$) and eastern (open
  circles, $150-205^{\circ}$) edges of the subcluster core.  An image
  showing the sectors used is also included (upper left).  The
  electron pressure profiles were produced by multiplying the
  deprojected electron density and the deprojected electron temperature
  profiles.}
\label{fig:corecomp}
\end{minipage}
\end{figure*}

\begin{figure}
\centering
\includegraphics[width=0.95\columnwidth]{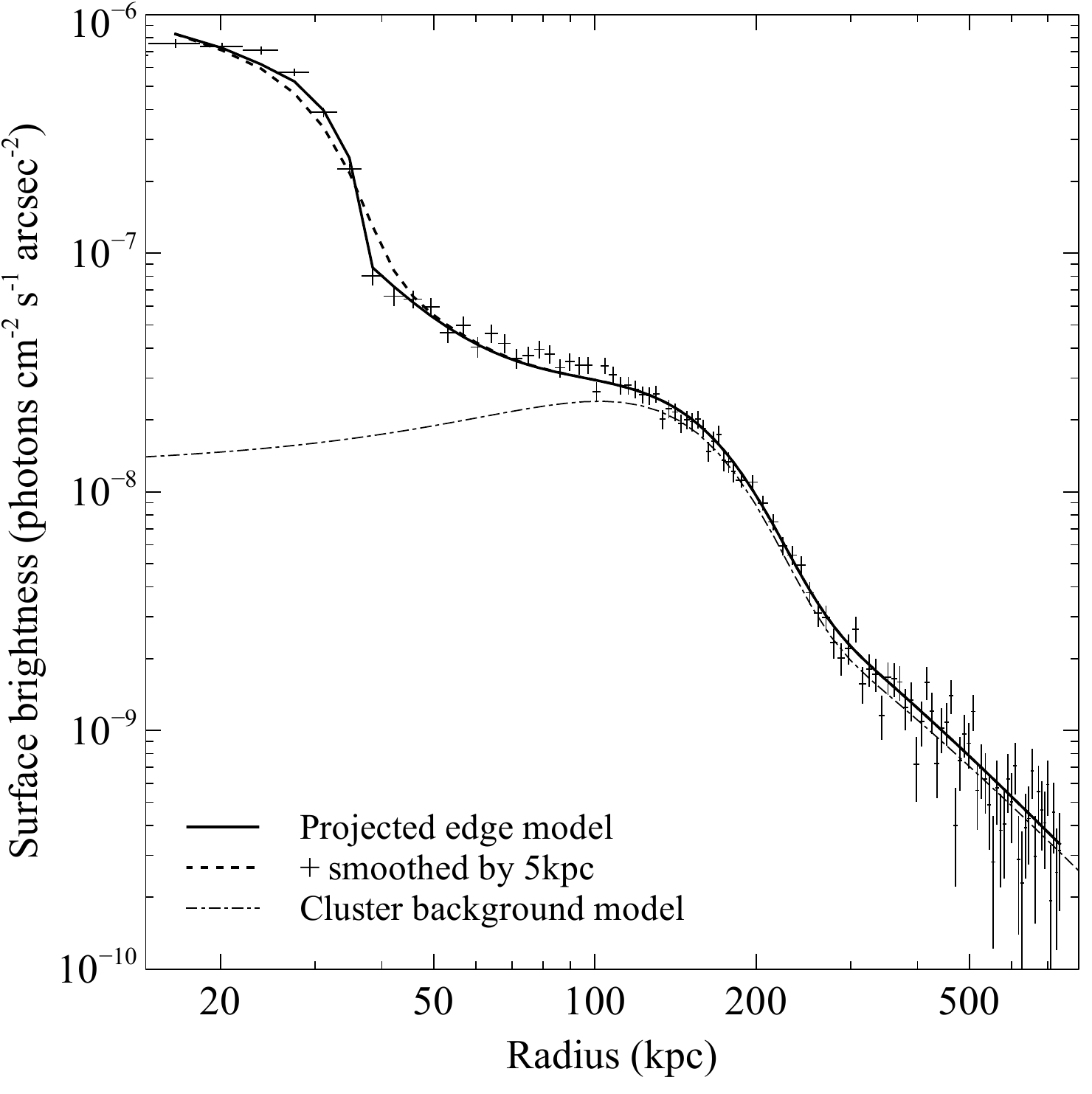}
\caption{Surface brightness profile for the southern edge of the subcluster
  core with the best-fit projected discontinuity model shown overlaid
  (solid line).  The dashed line corresponds to the best-fit projected
  discontinuity model smeared with a Gaussian of width $\sigma=5\kpc$
  and the dot-dashed line shows the model for the outer cluster
  emission.}
\label{fig:coreedge}
\end{figure}


In comparison, the SW edge of the subcluster core and tail appears to
be narrow and stable over a distance of $150\kpc$ (eg. Fig.
\ref{fig:mainimages}).  Fig. \ref{fig:corecomp} (upper right) shows that the
surface brightness drops by close to an order of magnitude in a
distance of only $\sim15\kpc$ across the narrowest point of the southern
edge.  The eastern edge of the subcluster core shows a much more gradual
decline in surface brightness with no obvious sharp edge corresponding
to a density jump.  The deprojected temperature, determined using
\textsc{projct} in \textsc{xspec}, increases rapidly across the narrow
southern edge of the subcluster core by a factor of $3.4^{+0.9}_{-0.6}$.  The
gas in the subcluster core is multi-phase and the temperature jump for
the cooler $0.5\keV$ component could be higher by a factor of a few.
Unless suppressed by at least an order of magnitude, thermal
conductivity in unmagnetized cluster gas should rapidly evaporate the
cool subcluster on a timescale of only $\sim10^7\yr$
(eg. \citealt{Ettori00}; \citealt{Markevitch03}; \citealt{Asai04}).

To determine the density jump across the subcluster southern edge, we fit the
surface brightness profile with the model for a projected spherical
density discontinuity discussed in section \ref{sec:shocks}.  A
surface brightness model for the outer cluster emission, a
$\beta$-model (\citealt{Cavaliere76}) plus a gaussian component to
account for the blurred bow shock edge, was added to the projected
discontinuity model to ensure the projected emission was accounted
for.  Fig. \ref{fig:coreedge} shows the best-fit model for the surface
brightness across the southern edge, assuming a discontinuous jump in the gas
density, which produced a best-fit density jump of $2.6\pm0.2$.  We
also smoothed the best-fit model by convolving the density jump with a
Gaussian function to determine if a non-zero width produced a better
fit to the surface brightness edge.  A model with $\sigma=0$ provided
the lowest $\chi^2$ and we calculated a 95\% upper limit on the cold
front edge width of $\sigma=2\kpc$.

This upper limit on the width can be compared with the Coulomb mean
free path of the electrons and protons on both sides of the cold
front.  As in section \ref{sec:shocks}, we estimate the mean free
path for particles crossing from inside the subcluster core to the
outside, $\lambda_{\mathrm{in{\rightarrow}out}}=0.5-1\kpc$, which is
the main source of diffusion across the edge.  If Coulomb diffusion is
not suppressed by any mechanism the southern edge should have a width of at
least several times $\lambda_{\mathrm{in{\rightarrow}out}}$.
Therefore it is likely that transport processes are highly suppressed
here.  As discussed in the introduction, the gas motion around the
subcluster core is likely to produce a preferentially tangential
magnetic field and strongly restrict heat flux and diffusion across
the front (eg. \citealt{VikhlininBfield01}; \citealt{Narayan01};
\citealt{Asai04,Asai05,Asai07}; \citealt{Lyutikov06};
\citealt{Xiang07}; \citealt{Dursi08}; see also \citealt{Churazov04}).


\subsection{Structure in the subcluster tail}


The NE and SW sides of the subcluster tail appear strikingly different
in this deeper \textit{Chandra} observation.  Even in the raw counts
image (Fig. \ref{fig:rawcounts}), the SW side of the tail exhibits a
sharp, coherent edge, that stretches $\sim150\kpc$ in length.
Fig. \ref{fig:mainimages} (lower left) indicates that there may be a
second fainter edge, $100\kpc$ in length, outside the main edge.  It's
unclear whether these are different or related structures as the
subcluster tail is clearly three-dimensional and structures are seen
in projection.  The narrow SW edge of the subcluster tail appears
similar to the sharp, straight edges of the subcluster bullet in the
Bullet cluster which appear unaffected by turbulence
(\citealt{Markevitch06}; \citealt{Markevitch07}).  Turbulence may be
suppressed along the SW edge of the core by a continuation of the
magnetic layer that strongly restricts conduction across the leading
edge of the cool core (section \ref{sec:conduction};
eg. \citealt{Lyutikov06}; \citealt{Xiang07}; \citealt{Dursi08}).

In comparison, the NW edge is broad ($\sim\mathrm{few}\times10\kpc$
across), and poorly defined as it is disrupted by the shearing flow of
the ambient ICM.  Small-scale instabilities are expected to grow
rapidly on timescales much shorter than the cluster passage time and
widen the interface.  If magnetic draping is responsible for the
stability of the SW edge of the subcluster core and tail, then it is
initially surprising that the eastern edge of the core is not similarly
stabilised.  However, as discussed in section \ref{sec:specradial},
the orbit of the subcluster core is likely to be curved to the South,
which will increase the relative velocity of the ambient ICM along the
NE edge of the core and reduce it along the SW edge.  The ambient gas
motion is responsible for the accumulation, stretching and ordering of
the magnetic fields along the subcluster interface and an alteration
in the subcluster trajectory could also affect the build up of this
magnetic draping layer.  There is no significant variation in the
thermal pressure along either side of the subcluster core
(Fig. \ref{fig:tailmaps}, left) but there could be a significant
variation in the flow velocity of the ambient ICM caused by the
subcluster trajectory, which is affecting the stability of the eastern edge.


Fig. \ref{fig:tailmaps} shows the variation in the ICM pressure and
temperature across the subcluster tail.  The thermal pressure peaks in
the dense, cool subcluster core but is also high behind the AGN where
the flow of the ambient ICM converges behind the core.  Although there
is some fluctuation in pressure through the subcluster tail, and the eastern
edge of the core appears to be breaking up, there are no significant
structures consistent with large-scale hydrodynamic instabilities
(eg. Abell 3667, \citealt{Mazzotta02}).  The cool blobs of gas
breaking off from the eastern edge of the cool core
(Fig. \ref{fig:coremaps}) appear to thermalize rapidly with the
surrounding hot ICM and there is a steadily increasing temperature
gradient through the subcluster tail.  There is some significant
variation in the temperature gradient along the northern edge of the
subcluster tail (Fig. \ref{fig:tailmaps}) where hotter $6-6.5\keV$ gas
appears to be impacting a region of cooler $5\keV$ material.  However,
it's not clear if this relatively cool region has been stripped from
the subcluster core and become isolated from the $\sim8\keV$ ambient
ICM or if it has a similar origin to the SW plume and is a remnant of
the primary cluster core.  Either way, this is likely to be a
transient feature that will thermalize with the surrounding ICM.

\section{Summary}
The deep \textit{Chandra} observations of Abell 2146 have revealed a
host of new and complex substructure.  We confirm the detection of
both a bow shock with Mach number $M=2.3\pm0.2$ and the first known
example of an upstream shock, which has a Mach number $M=1.6\pm0.1$.
The bow shock, located ahead of the cool subcluster core, can now be traced
to over $\sim500\kpc$ in length and appears significantly broader than
the upstream shock.  We find that the observed postshock electron
temperature profile behind the bow shock is lower than predicted for
instant shock-heating of the electrons and appears consistent with the
timescale for Coulomb collisional equilibration with the ions.  The
equivalent measurement for the upstream shock appears to support the
instant shock-heating model, similar to the result for the Bullet
cluster.  However, this measurement was more difficult for the
upstream shock as the Mach number was lower, the uncertainty on the
temperature values was greater and there is significant substructure
generated by the passage of the subcluster core.  We therefore place
less weight on the result from the upstream shock.  

In addition to the deviations from thermal equilibrium, we
considered the possibility of non-equilibrium ionization in the region
immediately behind each shock front.  However, the relatively low spectral
resolution, low photon count rates behind each shock front and large
radial bins compared to the ionization equilibration timescale make
this a difficult measurement with the existing data.  



Even in the raw counts image, the upstream shock appears remarkably
narrow and well-defined over $\sim440\kpc$ in length.  We calculate a
best-fit width for the shock of only $6^{+5}_{-3}\kpc$ (95\% errors),
which is significantly narrower than the estimated mean free path and
suggests that there is a significant suppression of Coulomb diffusion
across the shock front.  The bow shock appears broader with a best-fit
width of $12^{+6}_{-5}\kpc$ (95\% errors), however the measured width of the
shock fronts can be affected by deformations in the front shape which
smear the edge in projection.  Both the bow and upstream shock fronts
appear to be consistently narrow across their measured lengths.  


The deep \textit{Chandra} observation has also revealed a cool, dense
plume of material extending $\sim170\kpc$ in length in a direction
perpendicular to the merger axis.  This plume is likely to be the
remnant of the primary cluster core which has been pushed forward and
laterally (SW) by the impact with the subcluster core.  There does not
appear to be a symmetric feature extending to the NE which suggests
that the subcluster did not collide head on with the primary cluster.
If the subcluster passed to the northern side of the primary cluster,
simulations suggest that disrupted material from the primary core
would be mainly ejected in the direction of the observed plume.
However, a large impact parameter seems unlikely given the significant
disruption of the primary core and the symmetry of the two shock
fronts about the merger axis.


The surface brightness images of the subcluster core show a sharp,
leading edge corresponding to an increase in the gas density by a
factor of $2.6\pm0.2$ and temperature by a factor of at least
$3.4^{+0.9}_{-0.6}$.  The gas in the core has a metallicity of
$0.8\pm01\Zsun$ and is multiphase with significant temperature
components at $2\keV$, $1\keV$ and $0.5\keV$.  We find an upper limit
of $50\Msunpyr$ on the mass deposition rate in the subcluster core and
determine that $\sim80$\% of this cooling occurs in a bright filament.
The leading edge of the subcluster core is particularly narrow with a 95\% upper
limit on the width of only $2\kpc$, compared to the mean free path of
$0.5-1\kpc$, so Coulomb diffusion is significantly suppressed here.
We suggest that the motion of the subcluster core through the ambient
ICM has produced a magnetic draping layer which strongly restricts
conduction and diffusion across this edge.

The ram pressure stripped subcluster tail has interestingly different
structure along its two sides, which are both being sheared by the
surrounding medium.  Cool gas is being stripped off the eastern edge of the
core into a $45\kpc$ filament which has a steadily increasing
temperature along its length.  This material then breaks off the
filament into small blobs of gas which are directed into the
subcluster's wake.  The disrupted NE edge is broad at
$\sim\mathrm{few}\times10\kpc$ across.  In contrast, the SW edge
is narrow and well-defined over $\sim150\kpc$ in length.  The
magnetic draping layer around the leading edge of the core could be
stabilising the SW edge against turbulent instabilities but this
appears to be less effective along the NE edge.  For this non-zero
impact parameter merger, the trajectory of the subcluster is likely to
be curved to the southern, which will increase the velocity of the ambient
flow around the NE edge and could cause the observed gas stripping.

\section*{Acknowledgements}
HRR and BRM acknowledge generous financial support from the Canadian
Space Agency Space Science Enhancement Program.  HRR also acknowledges
support for this work provided by the National Aeronautics and Space
Administration through Chandra Award Number 16617775 issued by the
Chandra X-ray Observatory Center, which is operated by the Smithsonian
Astrophysical Observatory for and on behalf of the National
Aeronautics Space Administration under contract NAS8-03060.  ACF and LJK
thank the Royal Society for support.  REAC acknowledges funding from
the STFC.  SAB is supported in part by the Radcliffe Institute for
Advanced Study at Harvard University.  We thank the reviewer for helpful and constructive comments.  We thank Poshak Gandhi, Mike Irwin and
Sakurako Okamoto for help with the Subaru observations of Abell 2146.  We thank
Eugene Churazov, Roderick Johnstone and Julie Hlavacek-Larrondo for
helpful discussions.

\bibliographystyle{mn2e} 
\bibliography{refs.bib}

\begin{thebibliography}{86}
\expandafter\ifx\csname natexlab\endcsname\relax\def\natexlab#1{#1}\fi

\bibitem[{{Akahori} \& {Yoshikawa}(2010)}]{Akahori10}
{Akahori} T., {Yoshikawa} K., 2010, \pasj, 62, 335

\bibitem[{{Akahori} \& {Yoshikawa}(2011)}]{Akahori11}
{Akahori} T., {Yoshikawa} K., 2011, ArXiv e-prints

\bibitem[{{AMI Consortium: Rodr{\'{\i}}guez-Gonz{\'a}lvez}
  {et~al}\mbox{.}(2011){AMI Consortium: Rodr{\'{\i}}guez-Gonz{\'a}lvez},
  {Olamaie}, {Davies}, {Fabian}, {Feroz}, {Franzen}, {Grainge}, {Hobson},
  {Hurley-Walker}, {Lasenby}, {Pooley}, {Russell}, {Sanders}, {Saunders},
  {Scaife}, {Schammel}, {Scott}, {Shimwell}, {Titterington}, {Waldram}, \&
  {Zwart}}]{Rodriguez11}
{AMI Consortium: Rodr{\'{\i}}guez-Gonz{\'a}lvez} C. {et~al.}, 2011, \mnras,
  414, 3751

\bibitem[{{Arnaud}(1996)}]{Arnaud96}
{Arnaud} K.~A., 1996, in Astronomical Society of the Pacific Conference Series,
  Vol. 101, Astronomical Data Analysis Software and Systems V, {Jacoby} G.~H.,
  {Barnes} J., eds., pp. 17--+

\bibitem[{{Asai}, {Fukuda} \& {Matsumoto}(2004){Asai}, {Fukuda}, \&
  {Matsumoto}}]{Asai04}
{Asai} N., {Fukuda} N., {Matsumoto} R., 2004, Journal of Korean Astronomical
  Society, 37, 575

\bibitem[{{Asai}, {Fukuda} \& {Matsumoto}(2005){Asai}, {Fukuda}, \&
  {Matsumoto}}]{Asai05}
{Asai} N., {Fukuda} N., {Matsumoto} R., 2005, Advances in Space Research, 36,
  636

\bibitem[{{Asai}, {Fukuda} \& {Matsumoto}(2007){Asai}, {Fukuda}, \&
  {Matsumoto}}]{Asai07}
{Asai} N., {Fukuda} N., {Matsumoto} R., 2007, \apj, 663, 816

\bibitem[{{Balsara}, {Livio} \& {O'Dea}(1994){Balsara}, {Livio}, \&
  {O'Dea}}]{Balsara94}
{Balsara} D., {Livio} M., {O'Dea} C.~P., 1994, \apj, 437, 83

\bibitem[{{B{\"o}hringer} {et~al}\mbox{.}(2000){B{\"o}hringer}, {Voges},
  {Huchra}, {McLean}, {Giacconi}, {Rosati}, {Burg}, {Mader}, {Schuecker},
  {Simi{\c c}}, {Komossa}, {Reiprich}, {Retzlaff}, \&
  {Tr{\"u}mper}}]{Bohringer00}
{B{\"o}hringer} H. {et~al.}, 2000, \apjs, 129, 435

\bibitem[{{Borkowski}, {Lyerly} \& {Reynolds}(2001){Borkowski}, {Lyerly}, \&
  {Reynolds}}]{Borkowski01}
{Borkowski} K.~J., {Lyerly} W.~J., {Reynolds} S.~P., 2001, \apj, 548, 820

\bibitem[{{Brada{\v c}} {et~al}\mbox{.}(2006){Brada{\v c}}, {Clowe},
  {Gonzalez}, {Marshall}, {Forman}, {Jones}, {Markevitch}, {Randall},
  {Schrabback}, \& {Zaritsky}}]{Bradac06}
{Brada{\v c}} M. {et~al.}, 2006, \apj, 652, 937

\bibitem[{{Buote}(2000)}]{Buote00}
{Buote} D.~A., 2000, \mnras, 311, 176

\bibitem[{{Buote} \& {Canizares}(1994)}]{Buote94}
{Buote} D.~A., {Canizares} C.~R., 1994, \apj, 427, 86

\bibitem[{{Canning} {et~al}\mbox{.}(2011){Canning}, {Russell}, {Fabian},
  {Crawford}, \& {Hatch}}]{Canning11}
{Canning} R.~E.~A., {Russell} H.~R., {Fabian} A.~C., {Crawford} C.~S., {Hatch}
  N.~A., 2011, accepted to \mnras, astro-ph/1111.0452

\bibitem[{{Cash}(1979)}]{Cash79}
{Cash} W., 1979, \apj, 228, 939

\bibitem[{{Cavaliere} \& {Fusco-Femiano}(1976)}]{Cavaliere76}
{Cavaliere} A., {Fusco-Femiano} R., 1976, \aap, 49, 137

\bibitem[{{Churazov} \& {Inogamov}(2004)}]{Churazov04}
{Churazov} E., {Inogamov} N., 2004, \mnras, 350, L52

\bibitem[{{Clowe} {et~al}\mbox{.}(2006){Clowe}, {Brada{\v c}}, {Gonzalez},
  {Markevitch}, {Randall}, {Jones}, \& {Zaritsky}}]{Clowe06}
{Clowe} D., {Brada{\v c}} M., {Gonzalez} A.~H., {Markevitch} M., {Randall}
  S.~W., {Jones} C., {Zaritsky} D., 2006, \apjl, 648, L109

\bibitem[{{Clowe}, {Gonzalez} \& {Markevitch}(2004){Clowe}, {Gonzalez}, \&
  {Markevitch}}]{Clowe04}
{Clowe} D., {Gonzalez} A., {Markevitch} M., 2004, \apj, 604, 596

\bibitem[{{Dursi} \& {Pfrommer}(2008)}]{Dursi08}
{Dursi} L.~J., {Pfrommer} C., 2008, \apj, 677, 993

\bibitem[{{Ettori} \& {Fabian}(1998)}]{Ettori98}
{Ettori} S., {Fabian} A.~C., 1998, \mnras, 293, L33

\bibitem[{{Ettori} \& {Fabian}(2000)}]{Ettori00}
{Ettori} S., {Fabian} A.~C., 2000, \mnras, 317, L57

\bibitem[{{Fabian} \& {Daines}(1991)}]{Fabian91}
{Fabian} A.~C., {Daines} S.~J., 1991, \mnras, 252, 17P

\bibitem[{{Feretti} \& {Giovannini}(2008)}]{Feretti08}
{Feretti} L., {Giovannini} G., 2008, in Lecture Notes in Physics, Berlin
  Springer Verlag, Vol. 740, A Pan-Chromatic View of Clusters of Galaxies and
  the Large-Scale Structure, {M.~Plionis, O.~L{\'o}pez-Cruz, \& D.~Hughes},
  ed., pp. 143--+

\bibitem[{{Ferrari} {et~al}\mbox{.}(2008){Ferrari}, {Govoni}, {Schindler},
  {Bykov}, \& {Rephaeli}}]{Ferrari08}
{Ferrari} C., {Govoni} F., {Schindler} S., {Bykov} A.~M., {Rephaeli} Y., 2008,
  \ssr, 134, 93

\bibitem[{{Fox} \& {Loeb}(1997)}]{Fox97}
{Fox} D.~C., {Loeb} A., 1997, \apj, 491, 459

\bibitem[{{Freeman} {et~al}\mbox{.}(2002){Freeman}, {Kashyap}, {Rosner}, \&
  {Lamb}}]{Freeman02}
{Freeman} P.~E., {Kashyap} V., {Rosner} R., {Lamb} D.~Q., 2002, \apjs, 138, 185

\bibitem[{{Friedman} {et~al}\mbox{.}(1971){Friedman}, {Linson}, {Patrick}, \&
  {Petschek}}]{Friedman71}
{Friedman} H.~W., {Linson} L.~M., {Patrick} R.~M., {Petschek} H.~E., 1971,
  Annual Review of Fluid Mechanics, 3, 63

\bibitem[{{Fusco-Femiano}, {Landi} \& {Orlandini}(2005){Fusco-Femiano},
  {Landi}, \& {Orlandini}}]{Fusco-Femiano05}
{Fusco-Femiano} R., {Landi} R., {Orlandini} M., 2005, \apjl, 624, L69

\bibitem[{{Ghavamian}, {Laming} \& {Rakowski}(2007){Ghavamian}, {Laming}, \&
  {Rakowski}}]{Ghavamian07}
{Ghavamian} P., {Laming} J.~M., {Rakowski} C.~E., 2007, \apjl, 654, L69

\bibitem[{{G{\'o}mez} {et~al}\mbox{.}(2002){G{\'o}mez}, {Loken}, {Roettiger},
  \& {Burns}}]{Gomez02}
{G{\'o}mez} P.~L., {Loken} C., {Roettiger} K., {Burns} J.~O., 2002, \apj, 569,
  122

\bibitem[{{Hamilton}, {Sarazin} \& {Chevalier}(1983){Hamilton}, {Sarazin}, \&
  {Chevalier}}]{Hamilton83}
{Hamilton} A.~J.~S., {Sarazin} C.~L., {Chevalier} R.~A., 1983, \apjs, 51, 115

\bibitem[{{Heinz} {et~al}\mbox{.}(2003){Heinz}, {Churazov}, {Forman}, {Jones},
  \& {Briel}}]{Heinz03}
{Heinz} S., {Churazov} E., {Forman} W., {Jones} C., {Briel} U.~G., 2003,
  \mnras, 346, 13

\bibitem[{{Hull} {et~al}\mbox{.}(2001){Hull}, {Scudder}, {Larson}, \&
  {Lin}}]{Hull01}
{Hull} A.~J., {Scudder} J.~D., {Larson} D.~E., {Lin} R., 2001, \jgr, 106, 15711

\bibitem[{{Inogamov}(1999)}]{Inogamov99}
{Inogamov} N.~A., 1999, Astrophysics and Space Physics Reviews, 10, 1

\bibitem[{{Kalberla} {et~al}\mbox{.}(2005){Kalberla}, {Burton}, {Hartmann},
  {Arnal}, {Bajaja}, {Morras}, \& {P{\"o}ppel}}]{Kalberla05}
{Kalberla} P.~M.~W., {Burton} W.~B., {Hartmann} D., {Arnal} E.~M., {Bajaja} E.,
  {Morras} R., {P{\"o}ppel} W.~G.~L., 2005, \aap, 440, 775

\bibitem[{{Landau} \& {Lifshitz}(1959)}]{LandauLifshitz59}
{Landau} L.~D., {Lifshitz} E.~M., 1959, {Fluid mechanics}. Oxford, Pergamon
  Press

\bibitem[{{Lyutikov}(2006)}]{Lyutikov06}
{Lyutikov} M., 2006, \mnras, 373, 73

\bibitem[{{Macario} {et~al}\mbox{.}(2011){Macario}, {Markevitch},
  {Giacintucci}, {Brunetti}, {Venturi}, \& {Murray}}]{Macario11}
{Macario} G., {Markevitch} M., {Giacintucci} S., {Brunetti} G., {Venturi} T.,
  {Murray} S.~S., 2011, \apj, 728, 82

\bibitem[{{Markevitch}(2006)}]{Markevitch06}
{Markevitch} M., 2006, in ESA Special Publication, Vol. 604, The X-ray Universe
  2005, {A.~Wilson}, ed., pp. 723--+

\bibitem[{{Markevitch} {et~al}\mbox{.}(2002){Markevitch}, {Gonzalez}, {David},
  {Vikhlinin}, {Murray}, {Forman}, {Jones}, \& {Tucker}}]{Markevitch02}
{Markevitch} M., {Gonzalez} A.~H., {David} L., {Vikhlinin} A., {Murray} S.,
  {Forman} W., {Jones} C., {Tucker} W., 2002, \apjl, 567, L27

\bibitem[{{Markevitch} {et~al}\mbox{.}(2005){Markevitch}, {Govoni}, {Brunetti},
  \& {Jerius}}]{Markevitch05}
{Markevitch} M., {Govoni} F., {Brunetti} G., {Jerius} D., 2005, \apj, 627, 733

\bibitem[{{Markevitch} {et~al}\mbox{.}(2003){Markevitch}, {Mazzotta},
  {Vikhlinin}, {Burke}, {Butt}, {David}, {Donnelly}, {Forman}, {Harris}, {Kim},
  {Virani}, \& {Vrtilek}}]{Markevitch03}
{Markevitch} M. {et~al.}, 2003, \apjl, 586, L19

\bibitem[{{Markevitch} {et~al}\mbox{.}(2000){Markevitch}, {Ponman}, {Nulsen},
  {Bautz}, {Burke}, {David}, {Davis}, {Donnelly}, {Forman}, {Jones}, {Kaastra},
  {Kellogg}, {Kim}, {Kolodziejczak}, {Mazzotta}, {Pagliaro}, {Patel}, {Van
  Speybroeck}, {Vikhlinin}, {Vrtilek}, {Wise}, \& {Zhao}}]{Markevitch00}
{Markevitch} M. {et~al.}, 2000, \apj, 541, 542

\bibitem[{{Markevitch} \& {Vikhlinin}(2007)}]{Markevitch07}
{Markevitch} M., {Vikhlinin} A., 2007, \physrep, 443, 1

\bibitem[{{Mastropietro} \& {Burkert}(2008)}]{Mastropietro08}
{Mastropietro} C., {Burkert} A., 2008, \mnras, 389, 967

\bibitem[{{Mazzotta}, {Fusco-Femiano} \& {Vikhlinin}(2002){Mazzotta},
  {Fusco-Femiano}, \& {Vikhlinin}}]{Mazzotta02}
{Mazzotta} P., {Fusco-Femiano} R., {Vikhlinin} A., 2002, \apjl, 569, L31

\bibitem[{{McNamara} {et~al}\mbox{.}(2006){McNamara}, {Rafferty}, {B{\^i}rzan},
  {Steiner}, {Wise}, {Nulsen}, {Carilli}, {Ryan}, \& {Sharma}}]{McNamara06}
{McNamara} B.~R. {et~al.}, 2006, \apj, 648, 164

\bibitem[{{Montgomery}, {Asbridge} \& {Bame}(1970){Montgomery}, {Asbridge}, \&
  {Bame}}]{Montgomery70}
{Montgomery} M.~D., {Asbridge} J.~R., {Bame} S.~J., 1970, \jgr, 75, 1217

\bibitem[{{Murray} {et~al}\mbox{.}(1993){Murray}, {White}, {Blondin}, \&
  {Lin}}]{Murray93}
{Murray} S.~D., {White} S.~D.~M., {Blondin} J.~M., {Lin} D.~N.~C., 1993, \apj,
  407, 588

\bibitem[{{Mushotzky} \& {Szymkowiak}(1988)}]{Mushotzky88}
{Mushotzky} R.~F., {Szymkowiak} A.~E., 1988, in NATO ASIC Proc. 229: Cooling
  Flows in Clusters and Galaxies, {A.~C.~Fabian}, ed., pp. 53--62

\bibitem[{{Narayan} \& {Medvedev}(2001)}]{Narayan01}
{Narayan} R., {Medvedev} M.~V., 2001, \apjl, 562, L129

\bibitem[{{Ness}, {Scearce} \& {Seek}(1964){Ness}, {Scearce}, \&
  {Seek}}]{Ness64}
{Ness} N.~F., {Scearce} C.~S., {Seek} J.~B., 1964, \jgr, 69, 3531

\bibitem[{{Nulsen}(1982)}]{Nulsen82}
{Nulsen} P.~E.~J., 1982, \mnras, 198, 1007

\bibitem[{{O'Dea} {et~al}\mbox{.}(2008){O'Dea}, {Baum}, {Privon}, {Noel-Storr},
  {Quillen}, {Zufelt}, {Park}, {Edge}, {Russell}, {Fabian}, {Donahue},
  {Sarazin}, {McNamara}, {Bregman}, \& {Egami}}]{ODea08}
{O'Dea} C.~P. {et~al.}, 2008, \apj, 681, 1035

\bibitem[{{Owers} {et~al}\mbox{.}(2009){Owers}, {Nulsen}, {Couch}, \&
  {Markevitch}}]{Owers09}
{Owers} M.~S., {Nulsen} P.~E.~J., {Couch} W.~J., {Markevitch} M., 2009, \apj,
  704, 1349

\bibitem[{{Owers} {et~al}\mbox{.}(2011){Owers}, {Randall}, {Nulsen}, {Couch},
  {David}, \& {Kempner}}]{Owers11}
{Owers} M.~S., {Randall} S.~W., {Nulsen} P.~E.~J., {Couch} W.~J., {David}
  L.~P., {Kempner} J.~C., 2011, \apj, 728, 27

\bibitem[{{Poole} {et~al}\mbox{.}(2006){Poole}, {Fardal}, {Babul}, {McCarthy},
  {Quinn}, \& {Wadsley}}]{Poole06}
{Poole} G.~B., {Fardal} M.~A., {Babul} A., {McCarthy} I.~G., {Quinn} T.,
  {Wadsley} J., 2006, \mnras, 373, 881

\bibitem[{{Quillen} {et~al}\mbox{.}(2008){Quillen}, {Zufelt}, {Park}, {O'Dea},
  {Baum}, {Privon}, {Noel-Storr}, {Edge}, {Russell}, {Fabian}, {Donahue},
  {Bregman}, {McNamara}, \& {Sarazin}}]{Quillen08}
{Quillen} A.~C. {et~al.}, 2008, \apjs, 176, 39

\bibitem[{{Rafferty} {et~al}\mbox{.}(2006){Rafferty}, {McNamara}, {Nulsen}, \&
  {Wise}}]{Rafferty06}
{Rafferty} D.~A., {McNamara} B.~R., {Nulsen} P.~E.~J., {Wise} M.~W., 2006,
  \apj, 652, 216

\bibitem[{{Rakowski}(2005)}]{Rakowski05}
{Rakowski} C.~E., 2005, Advances in Space Research, 35, 1017

\bibitem[{{Randall} {et~al}\mbox{.}(2008){Randall}, {Markevitch}, {Clowe},
  {Gonzalez}, \& {Brada{\v c}}}]{Randall08}
{Randall} S.~W., {Markevitch} M., {Clowe} D., {Gonzalez} A.~H., {Brada{\v c}}
  M., 2008, \apj, 679, 1173

\bibitem[{{Raymond} \& {Korreck}(2005)}]{Raymond05}
{Raymond} J.~C., {Korreck} K.~E., 2005, in American Institute of Physics
  Conference Series, Vol. 781, The Physics of Collisionless Shocks: 4th Annual
  IGPP International Astrophysics Conference, {G.~Li, G.~P.~Zank, \&
  C.~T.~Russell}, ed., pp. 342--346

\bibitem[{{Rephaeli}, {Gruber} \& {Blanco}(1999){Rephaeli}, {Gruber}, \&
  {Blanco}}]{Rephaeli99}
{Rephaeli} Y., {Gruber} D., {Blanco} P., 1999, \apjl, 511, L21

\bibitem[{{Ricker} \& {Sarazin}(2001)}]{Ricker01}
{Ricker} P.~M., {Sarazin} C.~L., 2001, \apj, 561, 621

\bibitem[{{Roettiger}, {Stone} \& {Mushotzky}(1998){Roettiger}, {Stone}, \&
  {Mushotzky}}]{Roettiger98}
{Roettiger} K., {Stone} J.~M., {Mushotzky} R.~F., 1998, \apj, 493, 62

\bibitem[{{Russell}(2005)}]{Russell05}
{Russell} C.~T., 2005, in American Institute of Physics Conference Series, Vol.
  781, The Physics of Collisionless Shocks: 4th Annual IGPP International
  Astrophysics Conference, {G.~Li, G.~P.~Zank, \& C.~T.~Russell}, ed., pp.
  3--16

\bibitem[{{Russell} {et~al}\mbox{.}(2010){Russell}, {Sanders}, {Fabian},
  {Baum}, {Donahue}, {Edge}, {McNamara}, \& {O'Dea}}]{Russell10}
{Russell} H.~R., {Sanders} J.~S., {Fabian} A.~C., {Baum} S.~A., {Donahue} M.,
  {Edge} A.~C., {McNamara} B.~R., {O'Dea} C.~P., 2010, \mnras, 406, 1721

\bibitem[{{Sanders}(2006)}]{Sanders06}
{Sanders} J.~S., 2006, \mnras, 371, 829

\bibitem[{{Sanders} {et~al}\mbox{.}(2004){Sanders}, {Fabian}, {Allen}, \&
  {Schmidt}}]{Sanders04}
{Sanders} J.~S., {Fabian} A.~C., {Allen} S.~W., {Schmidt} R.~W., 2004, \mnras,
  349, 952

\bibitem[{{Sarazin}(1988)}]{Sarazin88}
{Sarazin} C.~L., 1988, {X-ray emission from clusters of galaxies}, {Sarazin,
  C.~L.}, ed.

\bibitem[{{Schekochihin} {et~al}\mbox{.}(2005){Schekochihin}, {Cowley},
  {Kulsrud}, {Hammett}, \& {Sharma}}]{Schekochihin05}
{Schekochihin} A.~A., {Cowley} S.~C., {Kulsrud} R.~M., {Hammett} G.~W.,
  {Sharma} P., 2005, \apj, 629, 139

\bibitem[{{Schekochihin} {et~al}\mbox{.}(2008){Schekochihin}, {Cowley},
  {Kulsrud}, {Rosin}, \& {Heinemann}}]{Schekochihin08}
{Schekochihin} A.~A., {Cowley} S.~C., {Kulsrud} R.~M., {Rosin} M.~S.,
  {Heinemann} T., 2008, Physical Review Letters, 100, 081301

\bibitem[{{Schwartz} {et~al}\mbox{.}(1988){Schwartz}, {Thomsen}, {Bame}, \&
  {Stansberry}}]{Schwartz88}
{Schwartz} S.~J., {Thomsen} M.~F., {Bame} S.~J., {Stansberry} J., 1988, \jgr,
  931, 12923

\bibitem[{{Smith} \& {Hughes}(2010)}]{Smith10}
{Smith} R.~K., {Hughes} J.~P., 2010, \apj, 718, 583

\bibitem[{{Spitzer}(1962)}]{Spitzer62}
{Spitzer} L., 1962, {Physics of Fully Ionized Gases}, {Spitzer, L.}, ed.

\bibitem[{{Springel} \& {Farrar}(2007)}]{Springel07}
{Springel} V., {Farrar} G.~R., 2007, \mnras, 380, 911

\bibitem[{{Struble} \& {Rood}(1999)}]{Struble99}
{Struble} M.~F., {Rood} H.~J., 1999, \apjs, 125, 35

\bibitem[{{Tidman} \& {Krall}(1971)}]{Tidman71}
{Tidman} D.~A., {Krall} N.~A., 1971, {Shock waves in collisionless plasmas},
  {Tidman, D.~A.~\& Krall, N.~A.}, ed.

\bibitem[{{Vikhlinin}, {Markevitch} \& {Murray}(2001{\natexlab{a}}){Vikhlinin},
  {Markevitch}, \& {Murray}}]{Vikhlinin01}
{Vikhlinin} A., {Markevitch} M., {Murray} S.~S., 2001{\natexlab{a}}, \apj, 551,
  160

\bibitem[{{Vikhlinin}, {Markevitch} \& {Murray}(2001{\natexlab{b}}){Vikhlinin},
  {Markevitch}, \& {Murray}}]{VikhlininBfield01}
{Vikhlinin} A., {Markevitch} M., {Murray} S.~S., 2001{\natexlab{b}}, \apjl,
  549, L47

\bibitem[{{Vikhlinin} \& {Markevitch}(2002)}]{Vikhlinin02}
{Vikhlinin} A.~A., {Markevitch} M.~L., 2002, Astronomy Letters, 28, 495

\bibitem[{{Wik} {et~al}\mbox{.}(2009){Wik}, {Sarazin}, {Finoguenov},
  {Matsushita}, {Nakazawa}, \& {Clarke}}]{Wik09}
{Wik} D.~R., {Sarazin} C.~L., {Finoguenov} A., {Matsushita} K., {Nakazawa} K.,
  {Clarke} T.~E., 2009, \apj, 696, 1700

\bibitem[{{Wong}, {Sarazin} \& {Ji}(2011){Wong}, {Sarazin}, \& {Ji}}]{Wong11}
{Wong} K.-W., {Sarazin} C.~L., {Ji} L., 2011, \apj, 727, 126

\bibitem[{{Xiang} {et~al}\mbox{.}(2007){Xiang}, {Churazov}, {Dolag},
  {Springel}, \& {Vikhlinin}}]{Xiang07}
{Xiang} F., {Churazov} E., {Dolag} K., {Springel} V., {Vikhlinin} A., 2007,
  \mnras, 379, 1325

\bibitem[{{Yoshikawa} \& {Sasaki}(2006)}]{Yoshikawa06}
{Yoshikawa} K., {Sasaki} S., 2006, \pasj, 58, 641

\end{thebibliography}

\clearpage

\end{document}